\documentclass[journal]{IEEEtran}

\usepackage{cite}
\usepackage{amsmath}
\usepackage{epsfig}
\usepackage{amssymb}
\usepackage{epstopdf}
\usepackage{amsthm}
\usepackage{amssymb}
\usepackage{graphicx}
\usepackage{color}
\newtheorem{lemma}{Lemma}
\newtheorem{proposition}{Proposition}
\newtheorem{theorem}{Theorem}
\newtheorem{remark}{Remark}
\newtheorem{coro}{Corollary}
\begin{document}

\title{Lattice-based Robust Distributed Source Coding}

\author{Dania~Elzouki, Sorina~Dumitrescu, and Jun Chen\thanks{The authors are with the Department of Electrical and Computer Engineering, McMaster University, Hamilton, Canada (Emails: elzoukda@mcmaster.ca; sorina/junchen@mail.ece.mcmaster.ca).}}
\maketitle

\begin{abstract}
In this paper, we propose a lattice-based robust distributed source coding system for two correlated sources and provide a detailed performance analysis under the high resolution assumption.
It is shown, among other things, that, in the asymptotic regime where 1) the side distortion approaches $0$ and 2) the ratio between the central and side distortions approaches $0$, our scheme is capable of achieving the  information-theoretic limit of quadratic multiple description coding when the two sources are identical, whereas a variant of the random coding scheme by Chen and Berger with Gaussian codes has a performance loss of 0.5 bits relative to this limit.
\end{abstract}


%
\IEEEpeerreviewmaketitle

\section{Introduction}

Distributed source coding is a crucial category of source coding problems, which has received significant attention over the past few decades. In distributed source coding, multiple correlated sources are encoded separately and sent to a central decoder for joint decoding. For the case when the central decoder is required to recover both sources losslessly, Slepian and Wolf \cite{Slepian} characterized the achievable rate region.
The case when  one source is available as side information at the decoder, while the other source may be recovered with some distortion,
was solved by Wyner and Ziv \cite{Wyner}. A general formulation of the distributed source coding problem in the lossy case was provided by Berger \cite{Berger} and Tung \cite{Tung}. However, the solution has been found only in certain special cases \cite{BergerYeung,Oohama,WTV08,WCW10,WC13,WC14}.

A closely related problem is the CEO problem introduced in \cite{CEO96}, where the correlated sources are noisy observations of a single remote source, whose reconstruction is required at the joint decoder. The rate-distortion region for this problem has been completely characterized in the quadratic Gausian case by Oohama \cite{Oohama2005} and Prabhakaran {\it et al.} \cite{Prab}.

Most of past work assume that the central decoder receives the information sent by all separate encoders. However, in practice this may not be true. For instance, in the case of wireless communications, the quality of the channels may be fluctuating. If the channel connecting some encoder with the fusion centre becomes very bad, the decoder is no longer able to recover the transmitted information. In such cases a robust system is desired. The robust version of the distributed source coding problem was considered in the CEO setting by Ishwar {\it al.} \cite{Ishwar} and Chen and Berger \cite{Chen}. The design of practical schemes was addressed in \cite{Saxena2006,Saxena2010,Xiaolin2016}, where iterative algorithms were employed for locally optimal designs. On the other hand, the work of Heegard and Berger \cite{Heegard} considers the robust version of the Wyner-Ziv problem and provides a characterization of the rate-distortion region.

The robust distributed source coding (RDSC) problem for the case of two correlated sources is considered in this paper. We propose a structured coding scheme based on lattices and provide a detailed performance analysis under the high resolution assumption. Note that when the two sources are identical, the setting being considered coincides with that of the classical multiple description coding (MDC) problem \cite{Ozarow,Wolf,EGC,Ahl,ZB,SLemma,WV07, WV09, Chen09, SSC14}.  For this case, our analysis indicates that, in the asymptotic regime where 1) the side distortion approaches $0$ and 2) the ratio between the central and side distortions approaches $0$, the proposed lattice-based scheme is capable of achieving the information-theoretic limit of quadratic MDC. For comparison we consider a variant of the random coding scheme originally proposed by Chen and Berger \cite{Chen} for the robust CEO problem and prove that the sum-rate of the latter system with Gaussian codes is $0.5$ bits higher than  the sum-rate of our proposed approach in the same asymptotic regime.

Our design is inspired by the prior work on multiple description lattice vector quantizers (MDLVQ) of Vaishampayan {\it et al.} \cite{MDLVQ} and Huang and Wu \cite{Huang}. It is worth pointing out that lattices have been used in prior work in other distributed source coding problems \cite{Zamir2002,Servetto,Pradhan2009,Merhav}. Most of the aforementioned papers use dithered lattice quantization, except for the work of Servetto \cite{Servetto}, which performs the analysis under the assumption of very high rate and very high correlation.

The  paper is structured as follows.  Section \ref{sec:pbForm}  presents the formulation of the RDSC problem. In Section \ref{sec:RCscheme} we analyze the performance of a random-coding-based RDSC scheme (similar to the one proposed in \cite{Chen}) with Gaussian codes and prove that it does not achieve the information-theoretic limit of quadratic MDC in the asymptotic regime where the side distortion and the ratio between the central and side distortions approach $0$.  Section \ref{sec:def} introduces definitions and notations related to lattices. Section \ref{main results} presents the main results of this work, namely the asymptotic performance analysis of the proposed lattice-based RDSC scheme. It is shown, among other things, that our design is able to achieve the fundamental limit of quadratic MDC in the aforementioned asymptotic regime. Section \ref{coding} presents the detailed operation of the  proposed lattice-based RDSC scheme. Finally, Section \ref{concl} concludes the paper.
%
%
%
%



\section{Problem Formulation}
\label{sec:pbForm}

\begin{figure}
\centering
\includegraphics[width=3.5in]{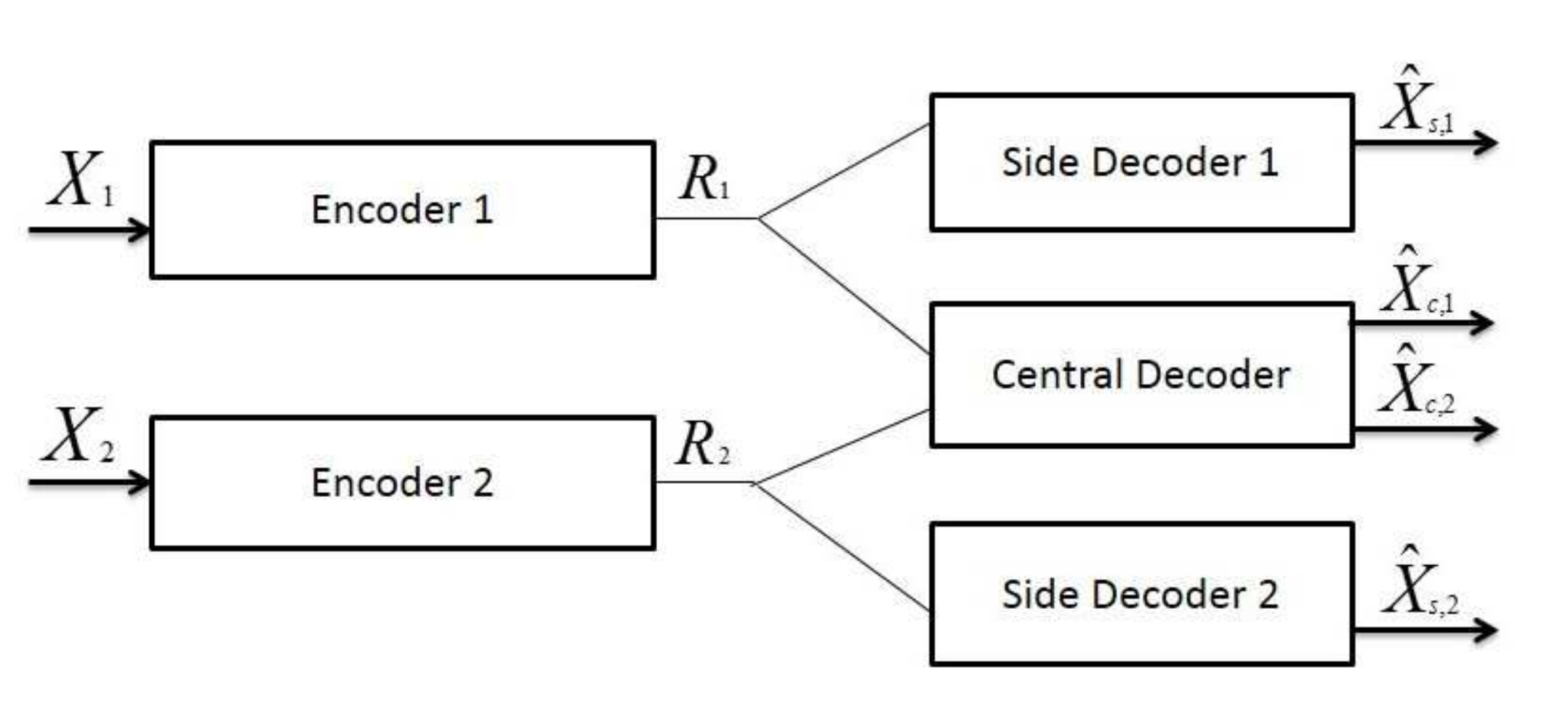}
\caption{Block diagram of robust distributed source coding.}
\label{RDSC}
\end{figure}

Consider two sources $X_1$ and $X_2$  with joint probability distribution  $f_{X_1X_2}$. The two sources generate a jointly i.i.d. random process ${(X_{1,k}, X_{2,k})}_{k\in \mathbb{N}}$.
We will consider an RDSC system as illustrated in Figure \ref{RDSC}. The system consists of two encoders and three decoders. Encoder $i$, $i=1,2$, has access only to source $X_i$, while the side decoder $i$ receives only the information sent by encoder $i$ and aims at reconstructing source $X_i$, $i=1,2$. The central decoder receives the information from both encoders and aims at reconstructing both sources $X_1$ and $X_2$.

For each $i=1,2$, let
$d_i:\mathcal{X}_i\times\hat{\mathcal{X}}_i\rightarrow[0,\infty)$  be a
distortion measure, where $\mathcal{X}_i$ and $\hat{\mathcal{X}}_i$ are
the source alphabet and the reconstruction alphabet for source $X_i$, respectively. The distortion measures are extended to sequences of length $n$ as follows
\begin{equation*}
d_i(x^n_i,\hat{x}^n_i)= \frac{1}{n}\sum\limits_{k=1}^n d_i(x_{i,k}, \hat{x}_{i,k}),
\end{equation*}
where $x_i^n=(x_{i,1}, \cdots, x_{i,n} )$, $\hat{x}_i^n=(\hat{x}_{i,1}, \cdots, \hat{x}_{i,n} )$.

A six-tuple $(R_{1},R_{2},d_{s,1},d_{s,2},d_{c,1}, d_{c,2})$ is said achievable,
if for any $\epsilon>0$ and all sufficiently large $n$, there
exist encoding functions
\begin{align*}
f^{(n)}_i:\mathcal{X}^n_i\rightarrow\{1,2,\cdots,\lfloor
2^{n(R_i+\epsilon)}\rfloor\}, \ i=1,2,
\end{align*}
and decoding functions
\begin{align*}
&g^{(n)}_{s,i}:\{1,2,\cdots,\lfloor2^{n(R_i+\epsilon)}\rfloor\}\rightarrow\hat{\mathcal{X}}_i^n,\ i=1,2,\\
&g^{(n)}_{c,i}:\{1,2,\cdots,\lfloor2^{n(R_1+\epsilon)}\rfloor\}\times\{1,2,\cdots,\lfloor 2^{n(R_2+\epsilon)}\rfloor\}\rightarrow\hat{\mathcal{X}}_i^n, \\
& \qquad \qquad \qquad \qquad \qquad \qquad \qquad \qquad \qquad \qquad i=1,2,
\end{align*}
such that
\begin{align*}
&\mathbb{E}\left[d_i(X^n_i,\hat{X}^n_{t,i})\right]\leq
d_{t,i}+\epsilon,\ i=1,2, \ t=s,c,
\end{align*}
where $\mathbb{E}[\cdot]$ denotes the expectation operator and
\begin{align*}
&\hat{X}_{t,i}^n=g^{(n)}_{t,i}(f^{(n)}_i(X^{n}_i)),\ i=1,2,\ t=s,c.
\end{align*}
The RDSC rate-distortion region, denoted by $\mathcal{RD}$, is the set of all such achievable six-tuples.

Furthermore, if $Y$ is a random variable over some discrete  alphabet $\mathcal{Y}$,  with probability mass function $p_Y$, and $\sum_{y\in\mathcal{Y}}p_Y(y)\log_2 p_Y(y)$ is finite, then the entropy of $Y$ is $H(Y)\triangleq -\sum_{y\in\mathcal{Y}}p_Y(y)\log_2 p_Y(y)$. If $X^n \in \mathbb{R}^n$ is a continuous random vector with probability density function (pdf) $f_{X^n}$, and the quantity $\int_{\mathbb{R}^n}f_{X^n}(x^n)\log_2 f_{X^n}(x^n) dx^n$ is finite, then the differential entropy of $X^n$ is $h(X^n)\triangleq -\int_{\mathbb{R}^n}f_{X^n}(x^n)\log_2 f_{X^n}(x^n) dx^n$.

\section{A Random-coding-based RDSC Scheme}
\label{sec:RCscheme}

In this section, we adapt a random coding scheme originally proposed by Chen and Berger \cite{Chen} for the robust CEO problem to the current setting and analyze the asymptotic performance of this scheme when specialized to the MDC scenario.

\begin{theorem}\label{thm:randomcoding}
We have $\mathcal{RD}_{in}\subseteq\mathcal{RD}$, where $\mathcal{RD}_{in}$ denotes the set of rate-distortion tuples $(R_1,R_2,d_{s,1},d_{s,2},d_{c,1},d_{c,2})$ for which there exist auxiliary random variables $U_{1},U_{2},W_{1},W_{2}$ (jointly distributed with the generic source variables $X_1$ and $X_2$) satisfying the following Markov chain
\begin{align}
W_1\leftrightarrow U_{1}\leftrightarrow X_{1}\leftrightarrow X_2\leftrightarrow U_2\leftrightarrow W_2,\label{eq:Markovcondition}
\end{align}
and deterministic mappings $g_{s,i}:\mathcal{W}_i\rightarrow\hat{\mathcal{X}}_i$, $g_{c,i}:\mathcal{U}_1\times\mathcal{U}_2\rightarrow\hat{\mathcal{X}}_i$, $i=1,2$, such that
\begin{align}
&R_{1}\geq I(X_{1};W_{1})+I(X_{1};U_{1}|U_{2},W_{1},W_{2}),\nonumber\\
&R_2\geq I(X_{2};W_{2})+I(X_{2};U_{2}|U_{1},W_{1},W_{2}),\nonumber\\
&R_{1}+R_{2}\geq I(X_{1};W_{1})+I(X_{2};W_{2}) \nonumber\\
& \qquad \qquad \qquad + I(X_{1},X_{2};U_{1},U_{2}|W_{1},W_{2}),\nonumber\\
&d_{s,i}\geq\mathbb{E}[d_i(X_i,g_{s,i}(W_i))],\quad i=1,2,\label{eq:distortionconstraint1}\\
&d_{c,i}\geq\mathbb{E}[d_i(X_i,g_{c,i}(U_1,U_2))],\quad i=1,2.\label{eq:distortionconstraint2}
\end{align}
\end{theorem}

The inner bound $\mathcal{RD}_{in}$ in Theorem \ref{thm:randomcoding} is achievable by the following random coding scheme. Roughly speaking, encoder $i$ produces $(W_i,U_i)$, where $W_i$ is a (lossy)
description of $X_i$, and $U_i$ is a refinement of $W_i$, $i=1,2$. Moreover, $W_i$ is encoded using the conventional lossy source code while $U_i$ is encoded using the Berger-Tung code \cite{Berger,Tung} with $(W_1,W_2)$ as the decoder side information, $i=1,2$. Side decoder $i$ can recover $W_i$ and use $g_{s,i}(W_i)$ as an estimate of $X_i$, $i=1,2$. The central decoder can recover $(U_1,U_2)$ (as well as $(W_1,W_2)$) and use $g_{c,i}(U_1,U_2)$ as an estimate of $X_i$, $i=1,2$.
The proof of Theorem \ref{thm:randomcoding} is similar to \cite[Theorem 1]{Chen} and is thus omitted.

In the rest of this paper, we assume $\mathcal{X}_1=\mathcal{X}_2=\hat{\mathcal{X}}_1=\hat{\mathcal{X}}_2=\mathbb{R}$ and adopt the squared distance as the distortion measure unless specified otherwise. To facilitate the evaluation of the achievable rate-distortion tuples in Theorem \ref{thm:randomcoding}, we shall focus on so-called Gaussian codes (in the sense of \cite{Zamir99}), which correspond to the following construction. Let
\begin{equation}
U_{i}=X_{i}+Z_{i},  W_{i}=U_{i}+Z_{i}^{'},\quad i=1,2,\label{eq:construction}
\end{equation}
where $Z_{1}, Z_{2}, Z_{1}^{'}, Z_{2}^{'}$ are zero-mean mutually independent Gaussian random variables and are independent of $(X_{1},X_{2})$. It is clear that $U_1,U_2,W_1,W_2$ constructed according to (\ref{eq:construction}) satisfy the Markov chain condition (\ref{eq:Markovcondition}). Moreover, we restrict $g_{s,i}$ and $g_{c,i}$, $i=1,2$, to be linear MMSE estimators; as such, (\ref{eq:distortionconstraint1}) and (\ref{eq:distortionconstraint2}) can be rewritten as
\begin{align}
&d_{s,i}\geq\mathrm{LMMSE}(X_i|W_i),\quad i=1,2,\label{eq:lmmse1}\\
&d_{c,i}\geq\mathrm{LMMSE}(X_i|U_1,U_2),\quad i=1,2,\label{eq:lmmse2}
\end{align}
where $\mathrm{LMMSE}$ denotes the squared distortion induced by the linear MMSE estimate.

Now consider the special case where $X_1=X_2=X$, $d_{s,1}=d_{s,2}=d_s$, and $d_{c,1}=d_{c,2}=d_{c}$. This is exactly the setting of the symmetric MDC problem. We shall assume that the source variable $X$ is of mean zero, variance $\sigma^2_X$, and finite differential entropy $h(X)$. It is well-known (see, e.g., \cite{Zamir99,chenGSO}) that in the asymptotic regime
\begin{align}
d_s\rightarrow 0, \frac{d_c}{d_s}\rightarrow 0,\label{eq:asyregime}
\end{align}
the minimum sum-rate of symmetric MDC is given by
\begin{align}
R_{MD}(d_s,d_c)=2h(X)-\frac{1}{2}\log_2(4(2\pi e)^2d_sd_c)+o(1).\label{eq:mdsumrate}
\end{align}
We shall show that in the same asymptotic regime the minimum sum-rate of the random-coding-based RDSC scheme in Theorem \ref{thm:randomcoding} with Gaussian codes as defined by (\ref{eq:construction})--(\ref{eq:lmmse2}) is given by
\begin{align}
R_{RC}(d_s,d_c)=2h(X)-\frac{1}{2}\log_2(2(2\pi e)^2d_sd_c)+o(1),\label{eq:rcsumrate}
\end{align}
therefore is 0.5 bits away from the fundamental limit.

First note that in the current setting (\ref{eq:lmmse1}) and (\ref{eq:lmmse2}) can be written equivalently as
\begin{align}
&d_{s}\geq\left(\frac{1}{\sigma^2_X}+\frac{1}{\sigma^2_{Z_i}+\sigma^2_{Z'_i}}\right)^{-1},\quad i=1,2,\label{eq:dcon1}\\
&d_c\geq\left(\frac{1}{\sigma^2_X}+\frac{1}{\sigma^2_{Z_1}}+\frac{1}{\sigma^2_{Z_2}}\right)^{-1},\label{eq:dcon2}
\end{align}
which implies
\begin{align}
&\sigma^2_{Z_i}+\sigma^2_{Z'_i}\leq(1+o(1))d_s,\quad i=1,2,\label{eq:imply1}\\
&\frac{\sigma^2_{Z_1}\sigma^2_{Z_2}}{\sigma^2_{Z_1}+\sigma^2_{Z_2}}\leq(1+o(1))d_c,\label{eq:imply2}
\end{align}
in the asymptotic regime (\ref{eq:asyregime}).
It can be verified that
\begin{align}
&I(X;W_{1})+I(X;W_{2})+I(X;U_{1},U_{2}|W_{1},W_{2})\nonumber\\
&=I(X;W_{1})+I(X;W_{2})+I(X;U_1,U_2)-I(X;W_1,W_2)\nonumber\\
&=h(W_1)-h(Z_1+Z'_1)+h(W_2)-h(Z_2+Z'_2)+ h(U_1,U_2)\nonumber\\
&\quad -h(Z_1,Z_2)-h(W_1,W_2)+h(Z_1+Z'_1,Z_2+Z'_2)\nonumber\\
&=h(W_1)+h(W_2)+h(U_1,U_2)-h(Z_1,Z_2)-h(W_1,W_2).\label{eq:tobesubinto}
\end{align}
We have
\begin{align}
& h(U_1,U_2)-h(W_1,W_2)\nonumber\\
&\quad =-I(Z'_1,Z'_2;X+Z_1+Z'_1,X+Z_2+Z'_2)\nonumber\\
&\quad =-I(Z'_1,Z'_2;Z_1+Z'_1-Z_2-Z'_2,X+Z_2+Z'_2)\nonumber\\
&\quad =-I(Z'_1,Z'_2;Z_1+Z'_1-Z_2-Z'_2)\nonumber\\
&\quad \quad-I(Z'_1,Z'_2;X+Z_2+Z'_2|Z_1+Z'_1-Z_2-Z'_2).\label{eq:tobesub}
\end{align}
Substituting (\ref{eq:tobesub}) into (\ref{eq:tobesubinto}) gives
\begin{align}
&I(X;W_{1})+I(X;W_{2})+I(X;U_{1},U_{2}|W_{1},W_{2})\nonumber\\
&=h(W_1)+h(W_2)-h(Z_1,Z_2)\nonumber\\
&\quad -I(Z'_1,Z'_2;Z_1+Z'_1-Z_2-Z'_2)\nonumber\\
&\quad-I(Z'_1,Z'_2;X+Z_2+Z'_2|Z_1+Z'_1-Z_2-Z'_2).\label{eq:tbc1}
\end{align}
Note that
\begin{align}
&h(Z_1,Z_2)+I(Z'_1,Z'_2;Z_1+Z'_1-Z_2-Z'_2)\nonumber\\
&=h(Z_1,Z_2)+h(Z_1+Z'_1-Z_2-Z'_2)-h(Z_1-Z_2)\nonumber\\
&=\frac{1}{2}\log_2\left(\frac{(2\pi e)^2\sigma^2_{Z_1}\sigma^2_{Z_2}(\sigma^2_{Z_1}+\sigma^2_{Z'_1}+\sigma^2_{Z_2}+\sigma^2_{Z'_2})}{\sigma^2_{Z_1}+\sigma^2_{Z_2}}\right)\nonumber\\
&\leq\frac{1}{2}\log_2\left(2(2\pi e)^2d_sd_c\right)+o(1)\label{eq:invoke1}
\end{align}
in the asymptotic regime (\ref{eq:asyregime}), where (\ref{eq:invoke1}) is due to (\ref{eq:imply1}) and (\ref{eq:imply2}). Moreover,
\begin{align}
&I(Z'_1,Z'_2;X+Z_2+Z'_2|Z_1+Z'_1-Z_2-Z'_2)\nonumber\\
&=h(X+Z_2+Z'_2|Z_1+Z'_1-Z_2-Z'_2)\nonumber\\
& \quad-h(X+Z_2+Z'_2|Z_1+Z'_1-Z_2-Z'_2,Z'_1,Z'_2)\nonumber\\
&=h(X+Z_2+Z'_2|Z_1+Z'_1-Z_2-Z'_2)\nonumber\\
&\quad-h(X+Z_2|Z_1-Z_2)\nonumber\\
&=h(X+\tilde{Z}_1)-h(X+\tilde{Z}_2),\label{eq:tbc2}
\end{align}
where $\tilde{Z}_1=Z_2+Z'_2-\mathbb{E}[Z_2+Z'_2|Z_1+Z'_1-Z_2-Z'_2]$ and $\tilde{Z}_2=Z_2-\mathbb{E}[Z_2|Z_1-Z_2]$. It can be shown \cite{TZ94} that in the asymptotic regime (\ref{eq:asyregime})
\begin{align*}
&h(W_i)=h(X)+o(1),\quad i=1,2,\\
&h(\tilde{Z}_i)=h(X)+o(1),\quad i=1,2,
\end{align*}
which together with (\ref{eq:tbc1}), (\ref{eq:invoke1}), and (\ref{eq:tbc2}) proves that
\begin{align*}
&I(X;W_{1})+I(X;W_{2})+I(X;U_{1},U_{2}|W_{1},W_{2})\nonumber\\
& \qquad \geq 2h(X)-\frac{1}{2}\log_2\left(2(2\pi e)^2d_sd_c\right)+o(1).
\end{align*}
The tightness of this lower bound can be established by choosing $\sigma^2_{Z_i}$, $\sigma^2_{Z'_i}$, $i=1,2$, that satisfy (\ref{eq:dcon1}) and (\ref{eq:dcon2}) with equalities.
This completes the proof of (\ref{eq:rcsumrate}).

There are two possible reasons why the performance of this random-coding-based RDSC scheme with Gaussian codes, when specialized to the symmetric MDC setting, is bounded away from the fundamental limit. Firstly, the restriction to Gaussian codes might be suboptimal. Secondly and more importantly, the random-coding-based RDSC scheme itself might be suboptimal. It is well known \cite{Ozarow, EGC} that the El Gamal-Cover (EGC) inner bound is tight for the quadratic Gaussian MDC problem. However, the inner bound $\mathcal{RD}_{in}$ in Theorem \ref{thm:randomcoding}, when specialized to the MDC setting, does not (at least expression-wise) coincide or subsume the EGC inner bound, therefore is unlikely to be tight. For the EGC inner bound, no Markov chain condition is imposed on the relevant auxiliary random variables. On the other hand, it is very difficult (if not impossible) to establish a single-letter inner bound of $\mathcal{RD}$ without a Markov chain condition similar to (\ref{eq:Markovcondition}). In other words, the conventional random coding argument seems to fall short of providing an RDSC scheme that does not have a performance gap when specialized to the MDC setting. This motivates us to develop an alternative RDSC scheme based on lattices that is able to close the gap in the MDC scenario.

\section{Lattice-related Definitions and Notations}
\label{sec:def}

Before introducing the proposed scheme we need to clarify the lattice-related definitions and notations to be used throughout this work, which is the purpose of this section.

We will denote by $x^n$ row vectors in $\mathbb{R}^n$. For $x^n=(x_1,\cdots,x_n)\in \mathbb{R}$ and $y^n=(y_1,\cdots,y_n) \in \mathbb{R}^n$, let $\langle x^n,y^n \rangle \triangleq \sum_{i=1}^n x_iy_i$,  and
$\| x^n \|\triangleq \sqrt{\langle x^n,x^n \rangle}$. We will use $\bf{0}$ for the all-zero $n$-dimensional vector. For any set $\mathcal{S}\subseteq \mathbb{R}^n$, any $\sigma \in \mathbb{R}$, and any $x^n\in \mathbb{R}^n$, denote
\begin{align*}
x^n + \mathcal{S} &\triangleq  \{x^n+y^n | y^n\in \mathcal{S}\},\\
\sigma \mathcal{S} &\triangleq \{\sigma y^n | y^n \in \Lambda\}.
\end{align*}
If $\mathcal{S}$ is a measurable set then $\nu(\mathcal{S})$ denotes its volume, i.e.,
\begin{equation*}
\nu(\mathcal{S}) \triangleq \int_{\mathcal{S}} d x^n.
\end{equation*}
An $n$-dimensional lattice $\Lambda$ is the set of all possible integer linear combinations of the rows of $\mathbf{G}$, for some  $n \times n$ non-singular matrix $\mathbf{G}$. In other words, we have
\begin{equation*}
\Lambda \triangleq \{\lambda \in \mathbb{R}^{n} |\lambda=i^n\cdot \mathbf{G}, i^n \in \mathbb{Z}^{n}\}.
\end{equation*}
The nearest-neighbor quantizer associated with the lattice $\Lambda$  is a function $Q_{\Lambda}(\cdot)$ which maps each $x^n \in \mathbb{R}^{n}$ to its nearest lattice point, i.e.,
\begin{equation}
\label{quantizer}
Q_{\Lambda}(x^n) \triangleq \arg \min_{\lambda \in \Lambda} \|x^n-\lambda \|.
\end{equation}
For every $\lambda \in \Lambda$ the set of all points mapped by $Q_{\Lambda}$ to $\lambda$ is the {\it Voronoi region $V_{\Lambda}(\lambda)$ of $\lambda$ in $\Lambda$}.
Note that the ties in (\ref{quantizer}) are broken in  a systematic manner such that
the following relation holds
\begin{equation*}
V_{\Lambda}(\lambda)=\lambda + V_{\Lambda}(0), \ \forall \lambda \in \Lambda.
\end{equation*}
For any set $\mathcal{S}\subseteq \mathbb{R}^n$, let $\overline{\mathcal{S}}$ denote the closure of the set $\mathcal{S}$, i.e., the union of $\mathcal{S}$ with its boundary. Then the following holds
\begin{equation*}
\overline{V_{\Lambda}(\lambda)}=\{x^n \in \mathbb{R}^n| \|x^n - \lambda\| \leq \|x^n - \lambda'\| \mbox{ for any }\lambda'\in \Lambda\}.
\end{equation*}
It is worth pointing out that, according to our definition of the Voronoi region, which follows \cite{Zamir}, not all the points on the boundary of $V_{\Lambda}(\lambda)$ are included in $V_{\Lambda}(\lambda)$, therefore $\overline{V_{\Lambda}(\lambda)}\neq V_{\Lambda}(\lambda)$.
We say that two Voronoi regions $V_{\Lambda}(\lambda_1)$ and $V_{\Lambda}(\lambda_2)$, where $\lambda_1, \lambda_2 \in \Lambda$, are {\it adjacent} if their closures have points in common.

Further, for any $x^n \in \mathbb{R}^{n}$ define
\begin{equation*}
x^n \hbox{ mod } \Lambda \triangleq  x^n - Q_{\Lambda}(x^n).
\end{equation*}
A {\it fundamental cell  of the lattice $\Lambda$} is a bounded set $\mathcal{C}_0$ which, when shifted by the lattice points, generates a partition of $\mathbb{R}^n$ \cite{Zamir}. In other words, the sets  $\lambda+\mathcal{C}_0$, for all $\lambda \in \Lambda$, form a partition  of $\mathbb{R}^n$.
All measurable fundamental cells of a lattice have the same volume \cite{Zamir}. This value is denoted by $\nu_{\Lambda}$ and we have $\nu_{\Lambda}=\nu(V_{\Lambda}(\bf{0}))$.
Further, for any set $\mathcal{S}\subset \mathbb{R}^n$, denote
\begin{equation*}
\bar{r}(\mathcal{S}) \triangleq \sup_{x^n\in \mathcal{S}} \|x^n\|.
\end{equation*}
The open ball of radius $r$ centered in the origin is denoted by $\mathcal{B}_{r}$, i.e.,
\begin{equation*}
\mathcal{B}_{r} \triangleq \{ x^n \in \mathbb{R}^{n} |\| x^n \| <r\}.
\end{equation*}
The {\it covering radius} of the lattice $\Lambda$ is $\bar{r}_{\Lambda} \triangleq \bar{r}(V_{\Lambda}(\bf{0}))$.
Additionally, we will denote by $r_{\Lambda}$ the {\it inscribed radius} of the lattice $\Lambda$, which is defined as the radius of the largest ball centered at the origin and included in $\overline{V_{\Lambda}(\bf{0})}$.

The {\it normalized second moment} of a measurable set $\mathcal{S}\subseteq \mathbb{R}^n$ is defined as
\begin{equation*}
G(\mathcal{S})\triangleq \frac{\int_{\mathcal{S}}\| x^n \| ^{2}dx}{n \nu(\mathcal{S})^{\frac{2}{n}+1}}.
\end{equation*}
It is important to notice that the normalized second moment is invariant to scaling.
The {\it normalized second moment of the lattice $\Lambda$}, denoted by $G_{\Lambda}$, is the normalized second moment of the Voronoi region of $\bf{0}$, i.e.,
\begin{equation*}
G_{\Lambda}\triangleq G(V_{\Lambda}(\bf{0})).
\end{equation*}
A pair of lattices $(\Lambda_{1}, \Lambda_{2})$ are said to be {\it nested} if $\Lambda_{2}\subset \Lambda_{1}$, i.e., if $\Lambda_{2}$ is a sublattice of $\Lambda_{1}$. The lattice $\Lambda_1$ is termed the {\it fine lattice}, while $\Lambda_2$ is termed the {\it coarse lattice}. The index of $\Lambda_2$ with respect to $\Lambda_1$ is $N(\Lambda_2:\Lambda_1)\triangleq \frac{\nu_{\Lambda_2}}{\nu_{\Lambda_1}}$. For any $\lambda_1\in \Lambda_1$, the set $\lambda_1+\Lambda_2$ is called a {\it coset of $\Lambda_2$ relative to $\Lambda_1$}. A set $\mathcal{F}\subset \Lambda_1$ is called a {\it set of coset representatives of $\Lambda_2$ relative to $\Lambda_1$} if the following two conditions hold
\begin{eqnarray*}
& \Lambda_1 = \cup_{\lambda_1\in \mathcal{F}} \left ( \lambda_1 + \Lambda_2 \right),\\
& \left ( \lambda_1 + \Lambda_2 \right) \cap \left ( \lambda'_1 + \Lambda_2 \right) = \emptyset \mbox{ for any } \lambda_1\neq \lambda'_1 \in \mathcal{F}.
\end{eqnarray*}
The above conditions imply that any point $\lambda\in \Lambda_1$ can be written in a unique way as $\lambda= \lambda_1 + \lambda_2$ where $\lambda_1\in \mathcal{F}$ and $\lambda_2\in \Lambda_2$. As shown in \cite{Zamir}, if $\mathcal{C}_0$ is a fundamental cell of the coarse lattice $\Lambda_2$, then the set $\mathcal{C}_0\cap \Lambda_1$ is a set of coset representatives of $\Lambda_2$ relative to $\Lambda_1$.

We use the squared error as  a distortion criterion. For any quantizer $Q$ defined on $\mathbb{R}^n$ and any random vector $X^n \in \mathbb{R}^n$ we denote by $D(Q,X^n)$ the per sample expected distortion, i.e.,
\begin{equation*}
D(Q,X^n)\triangleq \frac{1}{n}\mathbb{E}\left [\parallel Q(X^n)-X^n\parallel^2\right ].
\end{equation*}

\section{Main Results}
\label{main results}

As stated earlier in the paper, the main contribution of this work is the development of an RDSC scheme based on lattices, which is able to approach the theoretic performance limit of MDC in the asymptotic regime discussed in Section III. In this section we present the main results pertaining to the performance analysis of the  proposed scheme while the details of the scheme operation are deferred to the next section.

We will assume for the rest of the paper that the marginal pdfs $f_{X_1}$ and $f_{X_2}$  are continuous with finite marginal differential entropies $h(X_1)$ and $h(X_2)$. We additionally assume that $X_1$ and $X_2$ have mean zero and correlation coefficient $\rho$.

An $n$-dimensional lattice robust distributed source code (LRDSC, for short) operates on input sequences of length $n$ and is specified by a positive number $r_0$ and a
a triple of nested lattices in $\mathbb{R}^n$, $\mathcal{L}^{(n,r_0)}=(\Lambda_s, \Lambda_{in},\Lambda_c)$, where $\Lambda_s \subset  \Lambda_{in} \subset \Lambda_c$.
The finest lattice, $\Lambda_c$, called the {\it central} lattice, is used for the reconstruction at the central decoder. The coarsest lattice, $\Lambda_s$,  called the {\it side} lattice, is used for the reconstruction at the side decoders. The lattice $\Lambda_{in}$ is an auxiliary lattice used in the design; it is called the {\it intermediate} lattice, and is chosen such that the condition
\begin{equation}
\label{r_0}
r_0 +2\bar{r}_{c}\leq r_{in}
\end{equation}								
is satisfied, where $r_{in}$ denotes the inscribed radius of the lattice $\Lambda_{in}$, and $\bar{r}_{c}$ denotes the covering radius of the lattice $\Lambda_c$.
We point out that $\Lambda_s = \mu\Lambda_{in}$ for some even positive integer $\mu$.
The lattice $\Lambda_{s/2}\triangleq \frac{1}{2} \Lambda_s$,  called the {\it fractional} lattice, is also used in the operation of the scheme. Note that $\Lambda_s\subset \Lambda_{s/2}$. Since $\mu$ is an even number, we also have $\Lambda_{s/2}\subset \Lambda_{in}$.

The proposed LRDSC is designed such that when the input sequences $x_1^n$, $x_2^n$ are within distance $r_0$ from one another, the central decoder is able to refine the reconstruction of each source using the information received from the other encoder. On the other hand, when the above condition is violated,  the reconstruction at the central decoder has essentially the same quality as the reconstruction at the side decoder. For this reason
the probability
\begin{equation}
\label{pr0}
\mathcal{P}_{X_1X_2}(r_0)\triangleq \mathbb{P}[ X_{2}^{n}-X_{1}^{n} \notin {\mathcal{B}}_{r_{0}}]
\end{equation}
plays a crucial role in the performance of the scheme. As we will see shortly, the choice of $r_0$ governs the trade-off between the quality of the reconstruction at the central decoder and the encoder sum-rate.

In order to evaluate the performance of the LRDSC $\mathcal{L}^{(n,r_0)}$, we assume that there are $m$ consecutive sequences $x_i^n$ fed to each encoder $i$, one at a time. The outputs corresponding to all $m$ input sequences are further encoded losslessly. The rate and distortion of  the LRDSC $\mathcal{L}^{(n,r_0)}$ are defined in the limit of $m$ approaching $\infty$.
The notation $R(\mathcal{L}^{(n,r_0)})$ will be used for the sum-rate at the two encoders. Further,  the notations $d_{s,i}(\mathcal{L}^{(n,r_0)})$ and $d_{c,i}(\mathcal{L}^{(n,r_0)})$ are employed for the distortions of source $X_i$ at the side decoder $i$ and at the central decoder, respectively, for $i=1,2$. We will refer to $d_{s,i}(\mathcal{L}^{(n,r_0)})$ and $d_{c,i}(\mathcal{L}^{(n,r_0)})$ as the side distortion and the central distortion of source $i$, respectively, for $i=1,2$.

In order to simplify the notations related to the lattices involved in the scheme, we will use in the sequel only the subscript $c$, $in$, $s/2$, respectively $s$, instead of $\Lambda_c$, $\Lambda_{in}$, $\Lambda_{s/2}$, respectively $\Lambda_s$. For instance, we will use $\nu_c$ instead of $\nu_{\Lambda_c}$. Let us denote $K\triangleq N(\Lambda_{in}:\Lambda_c)=\frac{\nu_{in}}{\nu_c}$ and $M\triangleq N(\Lambda_s : \Lambda_{in})=\frac{\nu_{s}}{\nu_{in}}$. Since $\Lambda_s = \mu\Lambda_{in}$, it follows that $M=\mu^n$.

In this work, we evaluate the performance of the proposed lattice-based scheme  in the high resolution regime for fixed dimension $n$ unless stated otherwise.
More specifically, we require that the following relations hold simultaneously
\begin{eqnarray}
\label{lim0}
M\nu_s\rightarrow 0, \quad M\rightarrow \infty, \quad K \mbox{ is constant}.
\end{eqnarray}
Note that this asymptotic regime is similar in spirit to that considered in the prior work on MDLVQ \cite{MDLVQ,Huang,zhang}.
Clearly, the conditions specified in (\ref{lim0}) imply that $\nu_s, \nu_{in}$ and $\nu_c$ approach $0$. Further, since $\Lambda_{in}$ is a sublattice of $\Lambda_c$ such that
$r_0+2\bar{r}_c\leq r_{in}$, we also have that
\begin{eqnarray*}
r_0 = O(r_c) = O(\nu_c^{\frac{1}{n}})
\end{eqnarray*}
as (\ref{lim0}) holds.

In the formulations of the results in this section, we will use the statement that we have a family of LRDSCs satisfying (\ref{lim0}). This statement means that the family is parameterized by $\mu$ and $\theta>0$ and its members are the  LRDSCs $\mathcal{L}^{(n,r_0)}=(\Lambda_s,\Lambda_{in},\Lambda_c)$ satisfying
\begin{eqnarray*}
\Lambda_c=\theta\Lambda_{c,0}, \ \Lambda_{in}=\theta\Lambda_{in,0}, \ \Lambda_s=\mu\theta\Lambda_{in,0},
\end{eqnarray*}
for some fixed lattices $\Lambda_{in,0}\subset\Lambda_{c,0}$ in $\mathbb{R}^n$.
Then the asymptotic regime specified by (\ref{lim0}) is equivalently stated in terms of the parameters $\mu$ and $\theta$ as follows
\begin{eqnarray*}
\theta\rightarrow 0, \ \mu\rightarrow \infty, \ \mu^2\theta \rightarrow 0.
\end{eqnarray*}

Now we are ready to present the main result of this section.
\begin{theorem}
\label{th_main}
Consider a fixed pair of correlated sources $(X_1,X_2)$, a fixed positive integer $n$ and a family of LRDSCs $\mathcal{L}^{(n,r_0)}$ satisfying (\ref{lim0}). For $i=1,2$, let $U_i\triangleq Q_c(X_i^n) \mbox{ mod } \Lambda_{in}$. Then  in the asymptotic regime specified by (\ref{lim0}),
\begin{align}
\label{dsi}
& d_{s,i}(\mathcal{L}^{(n,r_0)})=\frac{1}{4}G_{s/2}(M\nu_s)^{\frac{2}{n}}(1+o(1)),\ i=1,2,\\
& G_c\nu_c^{\frac{2}{n}}(1+o(1))  \leq  d_{c,i}(\mathcal{L}^{(n,r_0)})  \leq \frac{1}{n}\kappa_0^2 \mathcal{P}_{X_1X_2}({r_0}) \left(M\nu_s\right)^{\frac{2}{n}} \nonumber \\
&  \qquad \qquad \qquad \qquad \qquad + G_c\nu_c^{\frac{2}{n}}(1+o(1)), \ i=1,2,\label{th3_eq012}\\
\label{R}
& R(\mathcal{L}^{(n,r_0)}) = h(X_1) + h(X_2)  - \frac{2}{n}\log_{2}\frac{\nu_s}{K^{1/2}} \nonumber \\
& \qquad \qquad \quad \   + \frac{1}{n}H(U_2|U_1) +o(1),
\end{align}
where $\kappa_0$ is a positive constant. Additionally, we have
\begin{equation}
\label{h(U2|U1)}
H(U_2|U_1)\leq  \log_2 K,
\end{equation}
while, if $r_0 \leq  r_c$,
\begin{align}
\label{cond_entropy}
& H(U_2|U_1)\leq 1 +
\left(1-\left(1-\frac{r_0}{r_c}\right)^n + \mathcal{P}_{X_1X_2}(r_0)\right)\log_2 K \nonumber \\
& \qquad \qquad \quad \ + o(1)
\end{align}
in the limit of  (\ref{lim0}). Furthermore, in each of relations (\ref{dsi})-(\ref{R}) and (\ref{cond_entropy}),  the term hidden in the little-o notation  can be upperbounded by a function which does not depend on the joint pdf $f_{X_1X_2}$ and approaches $0$ under  (\ref{lim0}).
\end{theorem}

The following corollary deals with the case when $\mathcal{P}_{X_1X_2}(r_0)$ is small enough to make the central distortion dominated by $G_c\nu_{c}^{\frac{2}{n}}$. In this case, the correlation coefficient between $X_1$ and $X_2$ must be close to $1$, therefore we will assume that the marginal pdfs are equal.
\begin{coro}
\label{coro:new1}
Consider a fixed pdf $f_{X}$, a fixed positive integer $n$ and a family of LRDSCs $\mathcal{L}^{(n,r_0)}$ satisfying (\ref{lim0}). Each LRDSC is applied to a pair of correlated sources $(X_1,X_2)$ with marginal pdfs equal to $f_{X}$, satisfying the condition
\begin{equation}
\label{pr0_cond1}
\mathcal{P}_{X_1X_2}(r_0) \leq \frac{\epsilon}{M^{\frac{4}{n}}},
\end{equation}
where $\lim_{(\ref{lim0})}\epsilon =0$. For $i=1,2$, let $U_i\triangleq Q_c(X_i^n) \mbox{ mod } \Lambda_{in}$.
Then in the limit of (\ref{lim0}),
\begin{align}
& d_{c,i}(\mathcal{L}^{(n,r_0)})= G_c\nu_{c}^{\frac{2}{n}}(1+o(1)),\ i=1,2,\label{final_dc1}\\
\label{rate_coro_new1}
& R(\mathcal{L}^{(n,r_0)})= 2h(X) + \frac{1}{2}\log_2 \frac{G_c G_{s/2}}{4d_{s,i}(\mathcal{L}^{(n,r_0)})d_{c,i}(\mathcal{L}^{(n,r_0)})}
\nonumber \\
& \qquad \qquad \quad \  + \frac{1}{n}H(U_2|U_1) + o(1), \ i=1,2.
\end{align}
If, additionally, we have $\lim_{(\ref{lim0})}\frac{r_0}{r_c}=0$, then
\begin{align}
&R(\mathcal{L}^{(n,r_0)})\nonumber\\
&= 2h(X) + \frac{1}{2}\log_2 \frac{G_c G_{s/2}}{4d_{s,i}(\mathcal{L}^{(n,r_0)})d_{c,i}(\mathcal{L}^{(n,r_0)})} + o(1).\label{rate_coro_new2}
\end{align}
Furthermore, in each of relations  (\ref{final_dc1})-(\ref{rate_coro_new2}) the term hidden in the little-o notation can be upperbounded by a function which depends on the joint pdf $f_{X_1X_2}$ only through $\mathcal{P}_{X_1X_2}(r_0)$ and approaches $0$ under (\ref{lim0}).
\end{coro}

\begin{remark}
\label{rem:corr}
Condition (\ref{pr0_cond1}) implies that, as the limits in (\ref{lim0}) are approached, the correlation coefficient between $X_1$ and $X_2$ approaches $1$, while the marginal distributions of $X_1$ and $X_2$ remain equal to the distribution of some random variable $X$. This raises the question  whether such a class of joint distributions exists. The answer is indeed positive as shown by the following argument. Construct first $X_1$ distributed as $X$. Then construct $X_0$ jointly distributed with $X_1$. Further, construct $X_2$ jointly distributed with $X_1$ and $X_0$ such that $X_1\leftrightarrow X_0 \leftrightarrow X_2$ form a Markov chain and the conditional distribution of $X_2$ given $X_0$ is the same as that of $X_1$ given $X_0$. It is clear that $X_1$ and $X_2$ have the same marginal distribution, and the correlation between $X_1$ and $X_2$ can be increased by appropriately increasing the correlation between $X_1$ and $X_0$.
\end{remark}

Let us assume now that the marginal pdfs of $X_1$ and $X_2$ are equal to the pdf of some random variable $X$ with variance  $\sigma_X^2$. We are interested in
finding a sufficient condition on the correlation coefficient $\rho$ under which relation (\ref{pr0_cond1}) holds.  To this end, we can apply  Markov's inequality to $\|X_2^n-X_1^n\|^2$, which leads to
\begin{align*} 
\mathcal{P}_{X_1X_2}({r_0})&=\mathbb{P}[ \|X_{2}^{n}-X_{1}^{n}\|^2 > {r_{0}^2}] \\
&<\frac{n\sigma^2_{X_2-X_1}}{r_0^2}\\
& =  \frac{2n(1-\rho)\sigma_X^2}{r_0^2}.
\end{align*}

By imposing further the condition
$\frac{n\sigma^2_{X_2-X_1}}{r_0^2} \leq \frac{\epsilon}{M^{\frac{4}{n}}}$
and using the fact that $r_0=O(\nu_{c}^{\frac{1}{n}})$, we obtain that
\begin{equation} \label{Markov1_var}
\sigma^2_{X_2-X_1} =  o\left(\frac{\nu_{c}^{\frac{2}{n}}}{M^{\frac{4}{n}}}\right), \mbox{ leading to } \rho = 1- o\left(\frac{\nu_{c}^{\frac{2}{n}}}{M^{\frac{4}{n}}}\right).
\end{equation}
This implies that $r_0$ can be chosen such that $r_0=o(\nu_{c}^{\frac{1}{n}})$, while  (\ref{pr0_cond1}) still holds.

On the other hand, for certain distributions Markov's inequality yields a loose bound, rendering the sufficient condition (\ref{Markov1_var}) too restrictive. This may happen if the distribution of the random variable $\|X_2^n-X_1^n\|^2$ has a light tail or a bounded support. For instance, if the random variable $|X_2-X_1|$ has as support the interval $[0,\frac{r_0}{\sqrt{n}}]$ for $r_0$ such that $\frac{r_0}{r_c}$ is constant under (\ref{lim0}), then we  can have $\sigma^2_{X_2-X_1}=\Theta(\nu_{c}^{\frac{2}{n}})$, while $\mathcal{P}_{X_1X_2}(r_0)=0$.

Next we will address the situation when (\ref{Markov1_var}) holds. In this case, according to Theorem \ref{th_main} and Corollary \ref{coro:new1}, we have
\begin{align*}
& d_{s,i}(\mathcal{L}^{(n,r_0)}) =  \frac{1}{4}G_{s/2}M^{\frac{4}{n}}K^{\frac{2}{n}}\nu_{c}^{\frac{2}{n}}(1+o(1)),\\
& d_{c,i}(\mathcal{L}^{(n,r_0)})= G_c\nu_{c}^{\frac{2}{n}}(1+o(1)).
\end{align*}
Then relation (\ref{Markov1_var}) is equivalent to $(1-\rho)\frac{d_{s,i}(\mathcal{L}^{(n,r_0)})}{d_{c,i}(\mathcal{L}^{(n,r_0)})^2}\rightarrow 0$, and further the limits in (\ref{lim0}) are
equivalent to
\begin{align}
&d_{s,i}(\mathcal{L}^{(n,r_0)})\rightarrow 0,  \quad\frac{d_{c,i}(\mathcal{L}^{(n,r_0)})}{d_{s,i}(\mathcal{L}^{(n,r_0)})}\rightarrow 0,  \nonumber\\
&(1-\rho)\frac{d_{s,i}(\mathcal{L}^{(n,r_0)})}{d_{c,i}(\mathcal{L}^{(n,r_0)})^2}\rightarrow 0.\label{cond_dist0}
\end{align}
Let us make the notations
\begin{equation}
\label{dsdc}
d_s=\frac{\sum_{i=1}^2d_{s,i}(\mathcal{L}^{(n,r_0)})}{2}, \
d_c=\frac{\sum_{i=1}^2d_{c,i}(\mathcal{L}^{(n,r_0)})}{2}.
\end{equation}
Then the limits in (\ref{cond_dist0}) imply that
\begin{equation}
\label{cond_dist}
d_{s}\rightarrow 0, \ \frac{d_{c}}{d_{s}}\rightarrow 0, \ (1-\rho)\frac{d_{s}}{d_{c}^2}\rightarrow 0.
\end{equation}
Let us denote by $R_L(n,d_{s},d_{c})$\footnote{This quantity is defined for those triples $(n,d_{s},d_{c})$ for which there exists an LRDSC $\mathcal{L}^{(n,r_0)}$ achieving average side distortion $d_s$ and average central distortion $d_{c}$.} the infimum of $R(\mathcal{L}^{(n,r_0)})$ over all $\mathcal{L}^{(n,r_0)}$ satisfying (\ref{dsdc}) for fixed $n$ and fixed pair $(X_1,X_2)$. Assume that the lattices used in the construction achieve the smallest second moment for the corresponding dimension, denoted by $G_{opt,n}$.
Applying this result in Corollary \ref{coro:new1}, we further obtain that
\begin{equation}
\label{prod21}
\left|R_L(n,d_{s},d_{c}) - 2h(X) - \frac{1}{2}\log_2 \frac{G^2_{opt,n}}{4d_{s}d_{c}} \right| \leq  \zeta(n,d_{s},d_{c},\rho),
\end{equation}
where $\lim_{(\ref{cond_dist})}\zeta(n,d_{s},d_{c},\rho)=0$.

Let us turn our attention to the case when $X_1=X_2=X$, i.e., $\rho=1$. In this case the asymptotic regime (\ref{cond_dist}) is specified only by $d_{s}\rightarrow 0$ and $\frac{d_{c}}{d_{s}}\rightarrow 0$. We will show that in this case our scheme achieves the fundamental limit of MDC. In order to formalize the result, we define the operational rate-distortion function of the proposed LRDSC in the case when $\rho=1$ as follows
\begin{equation*}
R_{L}(d_{s},d_{c}) \triangleq \inf_{n\geq 1} R_{L}(n,d_{s},d_{c}).
\end{equation*}
\begin{theorem}
\label{th_MD}
For any source $X$ with continuous pdf, the following holds
\begin{equation*}
\lim_{\stackrel{d_{s}\rightarrow 0}{\frac{d_{c}}{d_{s}}\rightarrow 0}}\left(R_{L}(d_{s},d_{c}) - R_{MD}(d_{s},d_{c})\right )=0,
\end{equation*}
where $R_{MD}(d_{s},d_{c})$ was defined in Section \ref{sec:RCscheme}.
\end{theorem}
\begin{IEEEproof}
Applying relation (\ref{prod21}) in the case of $\rho=1$ leads to
\begin{equation}
\label{prod21MD}
\lim_{\stackrel{d_{s}\rightarrow 0}{\frac{d_{c}}{d_{s}}\rightarrow 0}} \left(R_L(n,d_{s},d_{c}) - 2h(X) - \frac{1}{2}\log_2 \frac{G^2_{opt,n}}{4d_{s}d_{c}} \right) = 0.
\end{equation}
Using the fact that
$\lim\limits_{n \rightarrow \infty}G_{opt,n}=\frac{1}{2\pi e}$ \cite{zamir96} together with relation (\ref{eq:mdsumrate}) further leads to
\begin{equation}
\label{nogap}
\lim_{n\rightarrow \infty}\lim_{\stackrel{d_{s}\rightarrow 0}{\frac{d_{c}}{d_{s}}\rightarrow 0}}(R_L(n,d_{s},d_{c})- R_{MD}(d_{s},d_{c})) = 0.
\end{equation}
The above relation implies that for every $\epsilon>0$, there are  $n(\epsilon)$ and $\delta(\epsilon,n)$, for $n\geq n(\epsilon)$, such that
$R_{L}(n,d_{s},d_{c})-R_{MD}(d_{s},d_{c})<\epsilon$, for all $n\geq n(\epsilon)$ and $d_s\leq \delta(\epsilon,n), d_c/d_s\leq \delta(\epsilon,n)$.
Let $\xi(\epsilon)=\delta(\epsilon,n(\epsilon))$. Then whenever  $d_s\leq \xi(\epsilon)$ and $d_c/d_s\leq \xi(\epsilon)$, we have $R_{L}(d_{s},d_{c})-R_{MD}(d_{s},d_{c})\leq R_{L}(n(\epsilon),d_{s},d_{c})-R_{MD}(d_{s},d_{c})<\epsilon$. This implies that
\begin{equation*}
\lim_{\stackrel{d_{s}\rightarrow 0}{\frac{d_{c}}{d_{s}}\rightarrow 0}}\left(R_{L}(d_{s},d_{c}) - R_{MD}(d_{s},d_{c})\right )\leq 0.
\end{equation*}
Since the inequality $R_{L}(d_{s},d_{c}) - R_{MD}(d_{s},d_{c})\geq 0$ holds for all pairs $(d_s,d_c)$, the claim of the theorem follows.
\end{IEEEproof}

Note that another RDSC scheme which achieves the fundamental limit of MDC is a scheme which uses the encoders and decoders of an MDLVQ. Therefore, it is interesting to find out whether there is any advantage in using the proposed LRDSC scheme rather than directly applying an MDLVQ.

More specifically, in an  MDLVQ-based RDSC system, encoder $i$ maps the input sequence $x_i^n$ to $\lambda_{c,i} = Q_c(x_i^n)$,
next applies the index assignment $\alpha=(\alpha_1,\alpha_2):\Lambda_c \rightarrow \Lambda_s \times \Lambda_s$ and outputs the side lattice point $\alpha_i(\lambda_{c,i})$. Side decoder $i$ uses the received side lattice point $\lambda_{s,i}$ as the source reconstruction, while the central decoder looks for the central lattice point $\lambda_c$ satisfying $(\lambda_{s,1}, \lambda_{s,2})=(\alpha_1(\lambda_c), \alpha_2(\lambda_c))$, and uses $\lambda_c$ as the common  reconstruction for both sources.
The problem with this scheme is that, when $\lambda_{c,1} \neq \lambda_{c,2}$, the central distortion is essentially as high as the side distortion. To see this, note first that the mappings $\alpha_1, \alpha_2$ are constructed such that
$\alpha_1(\lambda_c') + \alpha_2(\lambda_c') = 2 Q_{s/2}(\lambda_c')$ for each $\lambda'_c \in \Lambda_c$.
Assume now that $Q_{s/2}(\lambda_{c,1}) = Q_{s/2}(\lambda_{c,2}) = \tau$ and $\lambda_{c,1} \neq \lambda_{c,2}$. Then $\alpha_1(\lambda_{c,1})\neq \alpha_1(\lambda_{c,2})$ because otherwise we would also have $\alpha_2(\lambda_{c,1})= \alpha_2(\lambda_{c,2})$, contradicting the fact that $\alpha$ is injective. Further, we obtain that $\alpha_1(\lambda_{c,1}) + \alpha_2(\lambda_{c,2}) \neq 2\tau$, which implies that the point $\lambda_c$ chosen by the central decoder is not in the same Voronoi region of the fractional lattice $\Lambda_{s/2}$ as $\lambda_{c,1}$ and $\lambda_{c,2}$. Then if $\|\lambda_{c,i}-\tau\| < 1/2 r_{s/2}$, the error in the reconstruction is at least $1/2 r_{s/2}$.
If $\sigma^2_{X_2-X_1} = \Theta(r_0^2)$ and $\frac{r_0}{r_c}$ is constant as the limits in (\ref{lim0})
are approached, the probability that $Q_{s/2}(\lambda_{c,1}) = Q_{s/2}(\lambda_{c,2}) = \tau$ and $\lambda_{c,1} \neq \lambda_{c,2}$ does not approach $0$, thus the central distortion cannot satisfy relation (\ref{final_dc1}), while the proposed RDSC scheme can.

\section{Detailed  Operation of the Proposed LRDSC Scheme}
\label{coding}
This section  presents in detail the operation of the proposed LRDSC.

\subsection{Preliminaries}

First we prove a key property, enabled by condition (\ref{r_0}),
which is essential in the operation of the LRDSC.
\begin{lemma}
\label{lemma:corr}
If $x_{2}^{n}- x_{1}^{n}\in {\mathcal{B}}_{r_0}$, then
\begin{align*}  \label{centraldis}
& \| Q_{c}(x_{1}^{n})-Q_{c}(x_{2}^{n}) \| < r_{in},\\
& \| Q_{in}(Q_{c}(x_{1}^{n}))-Q_{in}(Q_{c}(x_{2}^{n}))\|
< 3\bar{r}_{in}.
\end{align*}
\end{lemma}
\begin{IEEEproof}
Let $\lambda_{c,i}\triangleq Q_{c}(x_{i}^{n})$ and $\lambda_i \triangleq Q_{in}(\lambda_{c,i})$, for $i=1,2$.
Using the triangle inequality repeatedly, followed by (\ref{r_0}), one obtains that
\begin{align*}
\parallel \lambda_{c,1}- \lambda_{c,2} \parallel & \leq \parallel \lambda_{c,1}- x_1^n \parallel +
\parallel x_1^n - x_2^n \parallel + \parallel x_2^n- \lambda_{c,2} \parallel \\ & <
r_0 +2\bar{r}_{c}\leq r_{in}.
\end{align*}
Additionally,
\begin{align*} 
\parallel \lambda_{1}- \lambda_{2} \parallel & \leq \parallel \lambda_{1}- \lambda_{c,1} \parallel +
\parallel \lambda_{c,1}- \lambda_{c,2} \parallel + \parallel \lambda_{c,2}- \lambda_{2} \parallel \\ & <
r_{in} + 2\bar{r}_{in}  <  3\bar{r}_{in},
\end{align*}
which completes the proof.
\end{IEEEproof}

\begin{figure}
\centering
\includegraphics[scale=0.6]{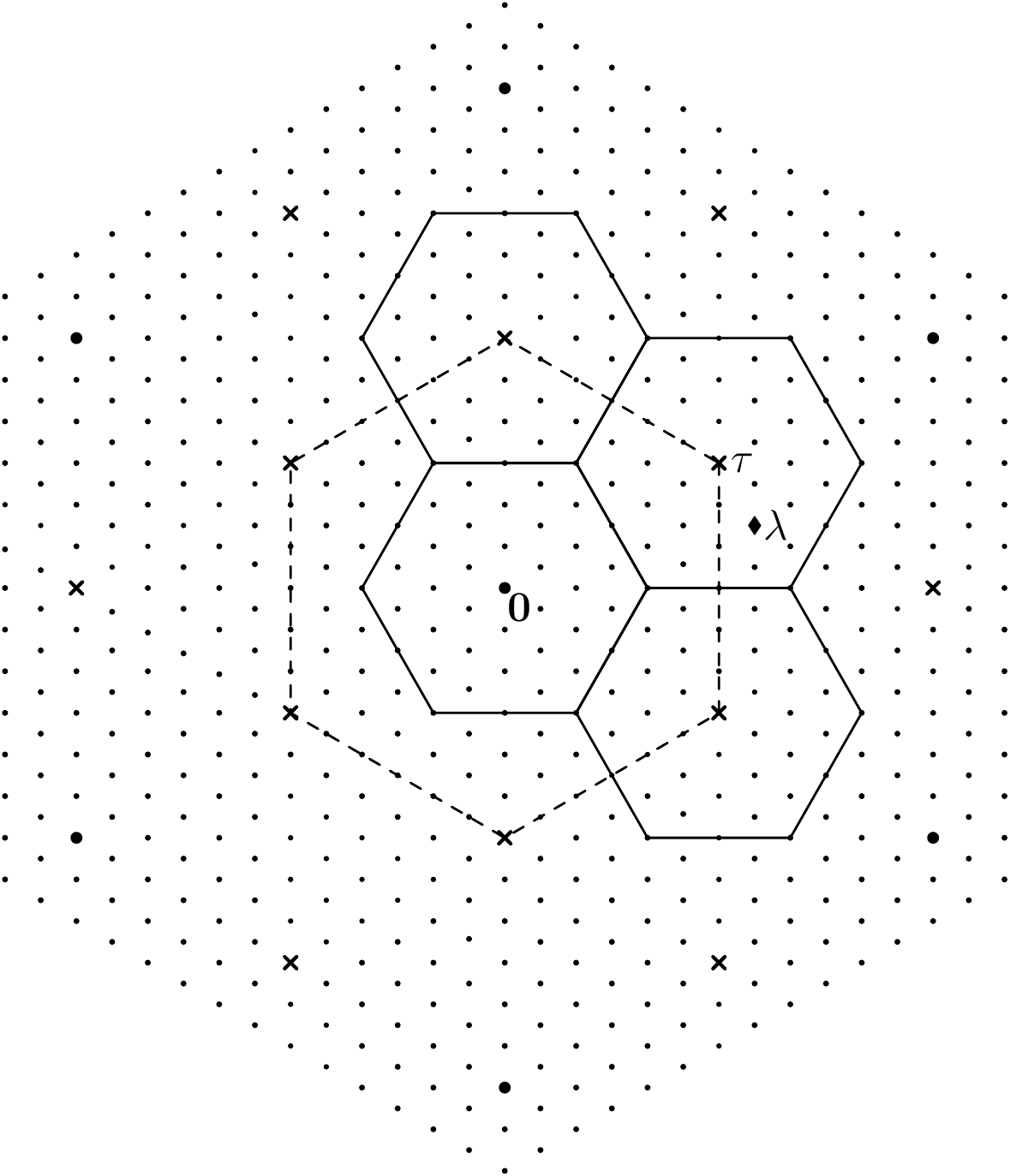}
\caption{Illustration of the lattices involved in the proposed scheme. The big dots represent points in $\Lambda_s$, while the small dots correspond to points in $\Lambda_{in}$. The value of $\mu$ is 12. The crosses represent the points in $\Lambda_{s/2}$ which are not in $\Lambda_s$. The hexagon drawn with dashed lines is the boundary of $V_s({\bf{0}})$. The centers of the four hexagons drawn with solid lines form the set $\mathcal{T}$. Each such hexagon is the boundary of a Voronoi region with respect to the lattice $\Lambda_{s/2}$. The point $\lambda$ represented by a diamond is an example of a point in the set $\mathcal{U}$, while $\tau=Q_{s/2}(\lambda)$.}
\label{fig:lattices}
\end{figure}					
      							
Next we define the labeling function $\beta_i:\Lambda_{in} \rightarrow \Lambda_s$ used at encoder $i=1,2$. For this we need to introduce some more notations as follows.   			
Let $\mathcal{T}\triangleq V_s({\bf{0}}) \cap \Lambda_{s/2}$. Then $\mathcal{T}$ is a set  of coset representatives of
$\Lambda_{s}$ relative to $\Lambda_{s/2}$. Thus, we have $|\mathcal{T}|=N(\Lambda_s:\Lambda_{s/2})=2^n$ and
\begin{equation*}
\Lambda_{s/2} =\bigcup_{\tau \in \mathcal{T}}\left(\tau + \Lambda_{s}\right).
\end{equation*}
It can be easily seen that the set $\cup_{\tau \in \mathcal{T}}V_{s/2}(\tau)$ is a fundamental cell of $\Lambda_s$. Denote $\mathcal{U}\triangleq \cup_{\tau \in \mathcal{T}}V_{s/2}(\tau)\cap \Lambda_{in}$. Then $\mathcal{U}$ is a set of coset representatives of $\Lambda_s$ relative to $\Lambda_{in}$, which implies that  $|\mathcal{U}|=N(\Lambda_s:\Lambda_{in})=M$ and
\begin{equation*}
\Lambda_{in} =\bigcup_{\lambda \in \mathcal{U}}\left(\lambda+ \Lambda_{s}\right).
\end{equation*}
We will first define $\beta_i$ for  $\lambda \in \mathcal{U}$ as follows
\begin{equation}
\label{def_bi_U}
\beta_{1}(\lambda) \triangleq \mu(\lambda -\tau ), \quad
\beta_{2}(\lambda) \triangleq 2\tau - \mu(\lambda -\tau ),
\end{equation}
where $\tau = Q_{s/2}(\lambda)$. Further, the mappings $\beta_1$ and $\beta_2$ are extended to $\Lambda_{in}$ using shifting. For arbitrary $\lambda \in \Lambda_{in}$, let $\lambda_{s/2}=Q_{s/2}(\lambda)$, i.e.,
$\lambda \in V_{s/2}(\lambda_{s/2})$. Then there is a unique pair
$(\tau,\lambda_s) \in \mathcal{T}\times \Lambda_s$ such that $\lambda_{s/2}=\lambda_s + \tau$. More specifically, we have $\lambda_s= Q_s(\lambda_{s/2})$ and $\tau= \lambda_{s/2}\mbox{ mod } \Lambda_s$. Then we define
\begin{align*}
&\beta_{1}(\lambda) \triangleq \beta_{1}(\lambda - \lambda_{s})+ \lambda_{s}= \mu(\lambda-\lambda_{s}-\tau) +\lambda_{s},\\
&\beta_{2}(\lambda) \triangleq \beta_{2}(\lambda - \lambda_{s}) + \lambda_{s}= 2\tau - \mu(\lambda-\lambda_{s}-\tau) + \lambda_{s}.
\end{align*}      			
The above definition implies that the mappings $\beta_i$ satisfy the {\it shift-invariance property}, i.e., that
\begin{equation*}
\beta_{i}(\lambda + \lambda'_s )= \beta_i(\lambda) + \lambda'_s, \quad \forall \lambda \in \Lambda_{in}, \quad \forall \lambda'_s \in \Lambda_s,\quad i=1,2.
\end{equation*}
The shift-invariance property further leads to the following relations, for $i=1,2$,
\begin{align}
\label{shiftbeta-1}
\beta_{i}^{-1}(\lambda_s ) & = \beta_i^{-1}({\bf{0}}) + \lambda_s, \forall \lambda_s \in \Lambda_s,\\
\label{shiftbeta-1of0}
\beta_{i}^{-1}({\bf{0}}) & = \{\lambda - \beta_i(\lambda)| \lambda \in \mathcal{U}\}.
\end{align}
Relation (\ref{shiftbeta-1}) is obvious. In order to prove (\ref{shiftbeta-1of0}), consider $\lambda'\in \Lambda_{in}$ and let
$(\lambda, \lambda_s) \in \mathcal{U} \times \Lambda_s$
be the unique pair such that $\lambda'=\lambda+ \lambda_s$. The shift-invariance property implies that $\beta_i(\lambda')= \beta_i(\lambda) + \lambda_s$, which leads to $\lambda_s =  \beta_i(\lambda') - \beta_i(\lambda)$.  Further, we obtain that $\lambda'= \lambda + \beta_i(\lambda') - \beta_i(\lambda)$. Consequently, the equality $\beta_i(\lambda')=0$ is equivalent to $\lambda'= \lambda - \beta_i(\lambda)$,  which proves the claim.

We point out that the construction of the mappings $\beta_1$ and $\beta_2$ was inspired by the index assignment used in MDLVQ \cite{MDLVQ,Huang} in two ways: 1) by defining the mappings on a set of coset representatives first and then extending them by shifting; 2) by imposing the condition that $\beta_1(\lambda)+\beta_2(\lambda)= 2Q_{s/2}(\lambda)$ for each $\lambda \in \Lambda_{in}$. On the other hand, it is important to note that we cannot simply use the mappings $\alpha_1,\alpha_2:\Lambda_{in}\rightarrow \Lambda_s$ that define the index assignment for MDLVQ\footnote{The lattice $\Lambda_{in}$ takes here the place of the central lattice on which the index assignment is defined for MDLVQ.} in \cite{MDLVQ,Huang} in place of our mappings $\beta_1$, $\beta_2$, since the requirement at the central decoder in our case is stronger than for MDLVQ. In particular, based on a received pair of side lattice points $\lambda_{s,1}$, $\lambda_{s,2}$, the central decoder of the MDLVQ uniquely identifies a  point $\lambda\in \Lambda_{in}$ such  that $(\alpha_1(\lambda),\alpha_2(\lambda))= (\lambda_{s,1},\lambda_{s,2})$. However, as we will see shortly, the central decoder in our scheme needs to uniquely identify two points $\lambda_1, \lambda_2 \in \Lambda_{in}$ such that $(\beta_1(\lambda_1),\beta_2(\lambda_2))= (\lambda_{s,1},\lambda_{s,2})$, using the additional knowledge of $\lambda_1-\lambda_2$. Using the pair of mappings $(\alpha_1,\alpha_2)$ designed for the MDLVQ in place of $(\beta_1,\beta_2)$ does not guarantee that the latter requirement is satisfied.

\subsection{LRDSC Operation}

\begin{figure*}
\centering
\includegraphics[width=0.8\textwidth,trim={1cm 7cm 5cm 3cm},clip]{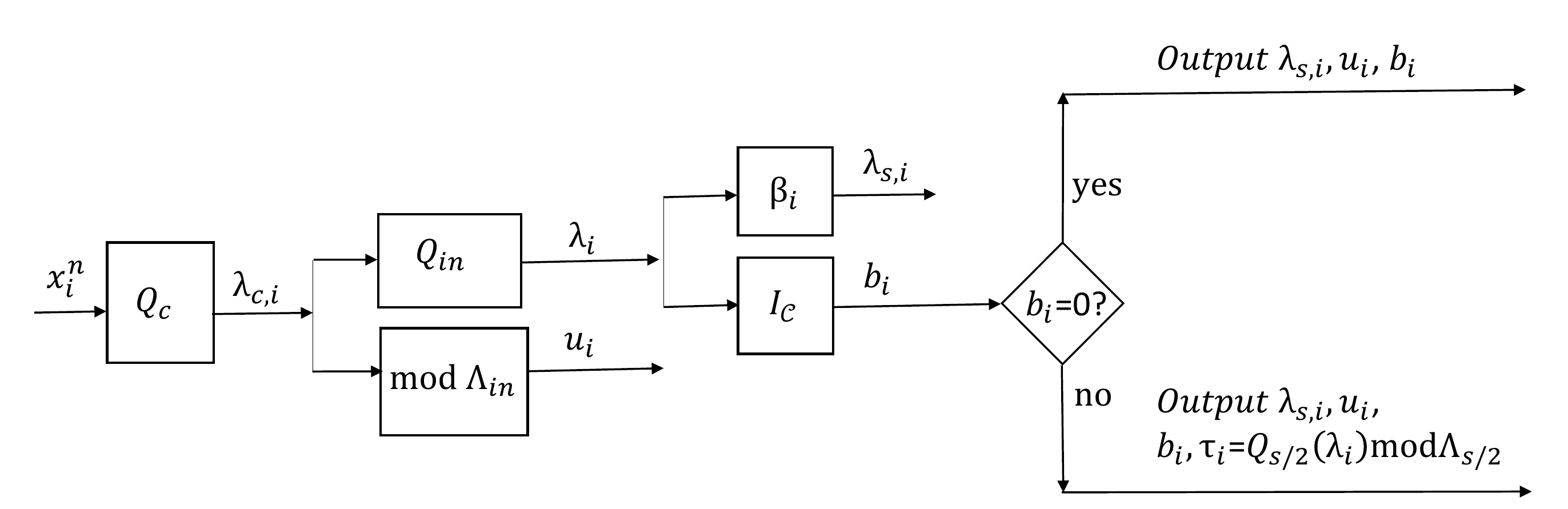}
\caption{Diagram describing the operation of encoder $i$, for $i=1,2$. $I_{\mathcal{C}}$ denotes the indicator function of the set $\mathcal{C}$.}
\label{fig:encoder}
\end{figure*}

Before describing the details of the proposed scheme we need the following discussion. Let us denote $\lambda_i=Q_{in}(Q_c(x_i^n))$, $i=1,2$.
Our scheme is designed such that side decoder $i$ will always be able to recover $\beta_i(\lambda_i)$, while the central decoder recovers $\lambda_{c,i}=Q_c(x_i^n)$, $i=1,2$, when the input sequences are sufficiently close, i.e., when $x_2^n-x_1^n \in \mathcal{B}_{r_0}$.  However, for the central decoder to achieve this goal, some additional information  needs to be transmitted besides $\beta_1(\lambda_1)$ and $\beta_2(\lambda_2)$. The amount of this additional information  is smaller when $\lambda_1$ and $\lambda_2$ are both in the same Voronoi cell of the lattice $\Lambda_{s/2}$.
Encoder $i$ is not able to determine all the time if this is the case or not, since it does not have knowledge of the other source sequence. However, based on Lemma \ref{lemma:corr}, if $\lambda_i\in V_{s/2}(\lambda_{s/2})$ and the distance from $\lambda_i$ to the boundary of $V_{s/2}(\lambda_{s/2})$ is not smaller than $3\bar{r}_{in}$, then encoder $i$ can infer that the other sequence is also in $V_{s/2}(\lambda_{s/2})$ when $x_2^n-x_1^n \in \mathcal{B}_{r_0}$.
Thus, we define the set
\begin{align}
\label{calC}
\mathcal{C}\triangleq \cup_{\lambda_{s/2} \in \Lambda_{s/2}} \mathcal{C}(\lambda_{s/2}),
\end{align}
where
\begin{align*}
\mathcal{C}(\lambda_{s/2})\triangleq V_{s/2}(\lambda_{s/2}) \setminus \left( \lambda_{s/2} + \gamma V_{s/2}({\bf{0}}) \right),
\end{align*}
for $\gamma\triangleq 1 - \frac{3\bar{r}_{in}}{r_{s/2}}$. According to Lemma \ref{lemma:corr}, if $\lambda_i \notin \mathcal{C}$, then $\lambda_{3-i}$ is in the same Voronoi cell of $\Lambda_{s/2}$ as $\lambda_i$, when $x_2^n-x_1^n \in \mathcal{B}_{r_0}$. Now we are ready to present the details of the encoder and decoder operation.

\noindent {\bf Encoder.}
Encoder $i$, for $i=1,2$, operates as follows (Fig. \ref{fig:encoder}). First the input sequence $x_{i}^{n}$ is quantized to the closest central lattice point $\lambda_{c,i}\triangleq Q_{c}(x_{i}^{n})$.
Next the point $\lambda_{c,i}$ is quantized to the closest point in the lattice $\Lambda_{in}$,   $\lambda_{i}\triangleq Q_{in}(\lambda_{c,i})$.   Let $u_i\triangleq\lambda_{c,i} \mbox{ mod }\Lambda_{in}$ and $\lambda_{s,i}\triangleq\beta_i(\lambda_i)$. Then encoder $i$ outputs $\lambda_{s,i}$,  $u_i$ and $b_i$, where $b_i=1$ if $\lambda_i \in \mathcal{C}$ and $b_i=0$ otherwise. Moreover, if $b_i=1$,  encoder $i$ also transmits $\tau_i\triangleq Q_{s/2}(\lambda_i) \mbox{ mod } \Lambda_s$. The first component, $\lambda_{s,i}$, will be used at the side decoder $i$, therefore, it is compressed using entropy coding before transmission. On the other hand, $u_1$ and $u_2$ are used only at the central decoder, therefore they will be compressed using Slepian-Wolf coding. Finally, $b_i$ and $\tau_i$ will also be used only at the central decoder, thus they may be compressed using Slepian-Wolf coding. However, we will use entropy coding to encode $b_i$ and fixed length codes for $\tau_i$ for simplicity of analysis, since, as shown in the proof of Theorem \ref{th_main}, the rate overhead is negligible asymptotically. Note that the aforementioned entropy coders and Slepian-Wolf coders are applied to blocks of $m$ symbols, where $m$ approaches $\infty$.\\

\noindent{\bf Decoder.}   	
Side decoder $i$, for $i=1,2$, outputs the reconstruction  $\hat{x}^n_{s,i} \triangleq \lambda_{s,i}$.
The central decoder recovers both values $\lambda_{s,1}$ and $\lambda_{s,2}$, and additionally, $u_1,u_2, b_1, b_2$.
First the decoder checks if the following condition is satisfied
\begin{eqnarray}
\label{dec_cond}
\|\lambda_{s,1}-\lambda_{s,2}\|  \leq  (8+c)\bar{r}_s + 3\bar{r}_{in}.
\end{eqnarray}
If the condition is violated, then the decoder concludes that $x_2^n-x_1^n\notin \mathcal{B}_{r_0}$, and outputs $\lambda_{s,i}$ as the reconstruction for source $i$, i.e., $\hat{x}^n_{c,i} \triangleq \lambda_{s,i}$, for $i=1,2$.

If condition (\ref{dec_cond}) is satisfied, the decoder assumes that $x_2^n-x_1^n\in \mathcal{B}_{r_0}$ and proceeds as follows.
First the following is computed
\begin{eqnarray}
\label{delta_lambda}
\tilde{\lambda} \triangleq Q_{in}(u_1-u_2).
\end{eqnarray}
Next the decoder proceeds based on the values of $b_1$ and $b_2$, and of $\tau_1$ and $\tau_2$ (if applicable), according to the following cases.
\begin{itemize}
\item[1)] If $b_1=0$ or $b_2=0$,  the decoder evaluates
\begin{align}
&\tilde{\lambda}_{s/2} \triangleq 1/2(\lambda_{s,1} + \lambda_{s,2}+\mu\tilde{\lambda}),\label{decoder_step21a} \\
&\tilde{\tau} \triangleq \tilde{\lambda}_{s/2} \mbox{ mod } \Lambda_s,\label{decoder_step21b}\\
&\tilde{\lambda}_1 \triangleq \tilde{\lambda}_{s/2} + \frac{1}{\mu}(\lambda_{s,1}  - \tilde{\lambda}_{s/2} + \tilde{\tau}),
\label{decoder_step22a}\\
&\tilde{\lambda}_2 \triangleq \tilde{\lambda}_{s/2} + \frac{1}{\mu}(\tilde{\tau} + \tilde{\lambda}_{s/2} - \lambda_{s,2}),\label{decoder_step22b}
\end{align}
and outputs the reconstructions $\hat{x}_{c,i}^n \triangleq \tilde{\lambda_i}+u_i$, for $i=1,2$.
\item[2)] If $b_1=b_2=1$ and $\tau_1=\tau_2$, the decoder proceeds as in case 1).
\item[3)] If $b_1=b_2=1$ and $\tau_1 \ne\tau_2$, then the decoder computes
\begin{align}
&\tilde{v} \triangleq 1/2(\lambda_{s,1} + \lambda_{s,2}+\mu\tilde{\lambda}-2\tau_2 - \mu(\tau_2-\tau_1)),
\label{decoder_step31aa}\\
&\hat{w} \triangleq \tilde{v} \mbox{ mod } \Lambda_s,\label{decoder_step31ab}\\
& \tilde{w} \triangleq \hat{w} - Q_s(\hat{w} +\frac{1}{2}(\tau_2-\tau_1)),\label{tilde_w}
\\
& \tilde{\lambda}_s \triangleq  \tilde{v} - (\mu+1) \tilde{w}, \quad \tilde{\lambda}'_s \triangleq  \tilde{\lambda}_s + 2\tilde{w}, \label{decoder_step31b}\\
&\tilde{\lambda}_1 \triangleq  \tilde{\lambda}_{s}+\tau_1 + \frac{1}{\mu}(\lambda_{s,1}  - \tilde{\lambda}_{s}), \label{decoder_step32a}\\
& \tilde{\lambda}_2 \triangleq \tilde{\lambda}'_{s} + \tau_2 + \frac{1}{\mu}(2\tau_2 + \tilde{\lambda}'_{s} - \lambda_{s,2}).\label{decoder_step32b}
\end{align}
Finally, the reconstructions are computed as $\hat{x}_{c,i}^n \triangleq \tilde{\lambda_i}+u_i$, for $i=1,2$. 		   						
\end{itemize}

\begin{proposition}
\label{prop_dc}
Let $\lambda_{c,i} \triangleq Q_{c}(x_{i}^{n})$, $\lambda_{i} \triangleq Q_{in}(\lambda_{c,i})$, $u_i \triangleq \lambda_{c,i} \mbox{ mod }\Lambda_{in}$, $\lambda_{s,i} \triangleq \beta_i(\lambda_i)$ and $\tau_i\triangleq Q_{s/2}(\lambda_i) \mbox{ mod } \Lambda_s$, for $i=1,2$. Then  when $x_2^n-x_1^n \in \mathcal{B}_{r_0}$ and the Slepian-Wolf decoding of $u_1$ and $u_2$ is successful, we have $\hat{x}^n_{c,i}=\lambda_{c,i}$, for $i=1,2$, and $\mu$ sufficiently large.
\end{proposition}
It is worth pointing out that a crucial aspect of the proposed scheme is the use of an identical quantizer for both sources (the quantizer defined by the central lattice). The use of a common finite-length quantizer in the two-source distributed coding scenario was  advocated earlier by Shirani and Pradhan in \cite{SP14} who argue that such a design preserves the correlation between sources more efficiently.

\section{Conclusion}
\label{concl}

We have proposed a constructive lattice-based scheme for robust distributed coding of two correlated sources. The analysis shows, among other things, that, in the asymptotic regime where 1) the side distortion approaches $0$ and 2) the ratio between the central and side distortions approaches $0$, our scheme is capable of approaching the information-theoretic limit of quadratic MDC when the two sources are identical, whereas a variant of the random-coding-based RDSC scheme by Chen and Berger with Gaussian codes is strictly sub-optimal. Note that in standard random coding arguments, to facilitate the joint typicality analysis, the block-length is often sent to $\infty$. However, in the infinite block-length limit, the condition needed to ensure joint typicality in the distributed setting is much more restrictive than its counterpart in the centralized setting; as a consequence, the resulting distributed coding schemes, when specialized to the centralized setting, may fail to achieve the fundamental performance limit. In contrast, for lattice-based schemes, the performance analysis can be carried out under fixed block-length (i.e., fixed lattice dimension), which reveals a smooth transition from the distributed setting to the centralized setting.
In this sense, our result echoes the recent finding in \cite{SP14} regarding the importance of finite block-length schemes in distributed source coding.


\appendices

\section{Proof of Proposition \ref{prop_dc}}
\label{proofProp1}

\begin{IEEEproof}[Proof of Proposition \ref{prop_dc}]
Assume that $x_2^n-x_1^n \in \mathcal{B}_{r_0}$ and that the Slepian-Wolf decoder employed at the central decoder is able to recover $u_1$ and $u_2$ correctly.
First we need to prove that condition (\ref{dec_cond}) is satisfied. To this end, we first show that
\begin{equation}
\label{bar_r}
\bar{r}(\beta^{-1}({\mathbf{0}})) \leq (4+\mu/2)\bar{r}_s.
\end{equation}
Note that relation (\ref{shiftbeta-1of0}) leads to
\begin{equation}
\label{bar_r_beta}
\bar{r}(\beta_i^{-1}({\bf{0}})) \leq \bar{r}(\mathcal{U}) + \bar{r}(\beta_i(\mathcal{U})).
\end{equation}
Further, since $\mathcal{T}\subset V_s({\bf{0}})$ and $V_{s/2}({\bf{0}})\subset V_s({\bf{0}})$, we obtain that $\mathcal{U}\subset \cup_{\tau \in \mathcal{T}}(\tau + V_{s/2}({\bf{0}})) \subset 2V_s({\bf{0}})$. Thus, $\bar{r}(\mathcal{U})\leq 2\bar{r}_s$. Moreover, from the definition of $\beta_i$ given in (\ref{def_bi_U}), we obtain that
$\bar{r}(\beta_i(\mathcal{U}))\leq 2\bar{r}(\mathcal{T}) + \mu\bar{r}_{s/2} \leq  2\bar{r}_s + \mu\bar{r}_{s/2} $. The above discussion, together with relation (\ref{bar_r_beta}) and the fact that $\bar{r}_{s/2}=1/2\bar{r}_s$, implies (\ref{bar_r}).

By applying the triangle inequality and the fact that $\|\lambda-\beta_i(\lambda)\|\leq \bar{r}(\beta^{-1}({\mathbf{0}}))$, together with Lemma \ref{lemma:corr}, we obtain
\begin{align*}
\|\lambda_{s,1}-\lambda_{s,2}\| & \leq \|\lambda_{s,1}-\lambda_1\| + \|\lambda_1-\lambda_2\| + \|\lambda_2-\lambda_{s,2}\| \\
& \leq 2\bar{r}(\beta^{-1}({\mathbf{0}})) + 3\bar{r}_{in}.
\end{align*}
Combining the above with (\ref{bar_r}) proves relation
(\ref{dec_cond}).

Using Lemma \ref{lemma:corr} and the fact that $\lambda_{c,i}= \lambda_i+u_i$, $i=1,2$,  we obtain that
\begin{eqnarray*}
r_{in} > \|\lambda_{c,1}-\lambda_{c,2}\|= \|u_1-u_2 - (\lambda_2-\lambda_1)\|,
\end{eqnarray*}
which, together with the fact that $\lambda_2-\lambda_1 \in \Lambda_{in}$, implies that $u_1-u_2\in V_{in}(\lambda_2-\lambda_1)$, i.e.,
$\lambda_2-\lambda_1=Q_{in}(u_1-u_2)$. This further implies that $\tilde{\lambda}$ computed in
(\ref{delta_lambda}) satisfies the equality
\begin{eqnarray}
\label{delta_lambda1}
\tilde{\lambda} = \lambda_2-\lambda_1.
\end{eqnarray}
Let $\lambda_s \triangleq Q_s(Q_{s/2}(\lambda_1))$  and  $\lambda'_s \triangleq Q_s(Q_{s/2}(\lambda_2))$. Using the fact that $\tau_i\triangleq Q_{s/2}(\lambda_i) \mbox{ mod } \Lambda_s$, for $i=1,2$, it follows that
$\lambda_1\in V_{s/2}(\lambda_s + \tau_1)$ and $\lambda_2\in V_{s/2}(\lambda'_s + \tau_2)$. Moreover, since $\lambda_{s,i}=\beta_i(\lambda_i)$ for $i=1,2$, we obtain that
\begin{align}
&\lambda_{s,1}= \mu(\lambda_1 -\lambda_s -\tau_1) + \lambda_s,\label{ls2a}\\
&\lambda_{s,2}= \lambda'_s + 2\tau_2 - \mu(\lambda_2 -\lambda'_s -\tau_2).\label{ls2b}
\end{align}
Assume now that case 1) holds. According to Lemma \ref{lemma:corr}, we have $\lambda_s + \tau_1 =\lambda'_s + \tau_2$. Since $\tau_1,\tau_2 \in \mathcal{T}$, it follows that $\lambda_s=\lambda'_s$ and $\tau_1=\tau_2$. Using further equations (\ref{decoder_step21a}), (\ref{decoder_step21b}), (\ref{delta_lambda1}), (\ref{ls2a}) and (\ref{ls2b}), we obtain that
$\tilde{\lambda}_{s/2}=\lambda_s +\tau_1$. This implies that $\tau_1=\tilde{\lambda}_{s/2}\mbox{ mod }\Lambda_s$, i.e., $\tilde{\tau}=\tau_1$. Equations (\ref{decoder_step22a}) and (\ref{decoder_step22b}) imply that $\tilde{\lambda}_i=\lambda_i$ and further that $\hat{x}_{c,i}^n=\lambda_{c,i}$, for $i=1,2$.

Assume now that $b_1=b_2=1$.
Let $\Delta_{s/2}$ denote the smallest distance between two points belonging, respectively, to the closures of two non-adjacent Voronoi regions of lattice $\Lambda_{s/2}$. When $\mu$ is large enough,
\begin{equation}
\label{cond_width}
\Delta_{s/2} > 3\bar{r}_{in}.
\end{equation}
Recall that, according to Lemma  \ref{lemma:corr}, we have $\|\lambda_1 - \lambda_2\| < 3 \bar{r}_{in}$. Condition (\ref{cond_width}) further ensures that
$\|\lambda_1 - \lambda_2\| < \Delta_{s/2}$, which implies that $V_{s/2}(\lambda_s+\tau_1)$ and $V_{s/2}(\lambda'_s + \tau_2)$ are either identical or adjacent.
Further, if $\tau_1=\tau_2$,
it follows that $\lambda'_s + \tau_2-(\lambda_s+\tau_1)\in \Lambda_s$. Thus, $V_{s/2}(\lambda_s+\tau_1)$ and $V_{s/2}(\lambda'_s + \tau_2)$ cannot be adjacent. Consequently, the equality $\lambda'_s + \tau_2 = \lambda_s+\tau_1$ holds and the proof proceeds as in case 1).

Assume now that $\tau_1\neq\tau_2$.
Then $\lambda_s+\tau_1\neq \lambda'_s + \tau_2$.
Denote
$\Delta\lambda_{s/2}\triangleq \lambda'_s + \tau_2-(\lambda_s+\tau_1)$.
Then ${\bf{0}}$ and $\Delta\lambda_{s/2}$ are adjacent points of the lattice $\Lambda_{s/2}$ (i.e., their Voronoi regions are adjacent). It follows that
\begin{equation}
\label{rel_delta}
\Delta\lambda_{s/2}\in \overline{V_s({\bf{0}})}.
\end{equation}
Let $w\triangleq \frac{1}{2}(\lambda'_s-\lambda_s)$.
Using equations (\ref{decoder_step31aa}), (\ref{decoder_step31ab}), (\ref{delta_lambda1}), (\ref{ls2a}) and (\ref{ls2b}), we obtain that
\begin{eqnarray}
\label{eqtildev}
\tilde{v} = \lambda_s +\frac{\mu}{2}(\lambda'_s-\lambda_s)+w.
\end{eqnarray}
Since $\mu$ is even, it follows that $\frac{\mu}{2}(\lambda'_s-\lambda_s)\in \Lambda_s$. Thus, $w\mbox{ mod }\Lambda_s = \tilde{v}\mbox{ mod }\Lambda_s = \hat{w}$. It follows that $w = \bar{\lambda}_s + \hat{w}$ for some $\bar{\lambda}_s\in \Lambda_s$. Then $\Delta\lambda_{s/2}=2w + \tau_2-\tau_1=2(\bar{\lambda}_s + \hat{w}) + \tau_2-\tau_1$. Using further (\ref{rel_delta}) leads to
$\frac{1}{2}\Delta\lambda_{s/2}=\bar{\lambda}_s + \hat{w} + \frac{1}{2}(\tau_2-\tau_1)\in \frac{1}{2}\overline{V_s({\bf{0}})}\subset V_s({\bf{0}})$, which further implies that $ -\bar{\lambda}_s=Q_s(\hat{w} + \frac{1}{2}(\tau_2-\tau_1))$.
It follows that $\tilde{w}=w$, where $\tilde{w}$ is defined in (\ref{tilde_w}). Combining this with (\ref{decoder_step31b}) and (\ref{eqtildev}), we obtain that $\tilde{\lambda}_s=\lambda_s$ and
$\tilde{\lambda}'_s=\lambda'_s$. Finally, equations (\ref{decoder_step32a}) and (\ref{decoder_step32b}) imply that $\tilde{\lambda}_i=\lambda_i$ and further that $\hat{x}_{c,i}^n=\lambda_{c,i}$, for $i=1,2$.
\end{IEEEproof}

\section{Proof of Results in Section \ref{main results}}
\label{proofth}

Before proceeding to the proof of Theorem \ref{th_main}, we need a  few more notations and some auxiliary results.

Consider an LRDSC $\mathcal{L}^{(n,r_0)}=(\Lambda_s,\Lambda_{in},\Lambda_c)$.
For each $\lambda_s \in \Lambda_s$ and $i=1,2$, let $\mathcal{A}_i(\lambda_s)\triangleq \{x_i^n|  \hat{x}_{s,i}^n=\lambda_s\}$. Further,
for each $\lambda\in \Lambda_{in}$, denote $\mathcal{M}(\lambda)\triangleq \cup_{\lambda_c \in V_{in}(\lambda)\cap \Lambda_c} V_c(\lambda_c)$.
Then
$\mathcal{A}_i(\lambda_s)= \cup_{\lambda \in \beta_{i}^{-1}(\lambda_{s})} \mathcal{M}(\lambda)$.    Clearly, we have $\mathcal{M}(\lambda)=\lambda + \mathcal{M}({\bf{0}})$ for all $\lambda \in \Lambda$. This fact, together with relation (\ref{shiftbeta-1}), implies that
\begin{equation}
\mathcal{A}_i(\lambda_s) = \mathcal{A}_i({\bf{0}}) + \lambda_s, \quad \forall \lambda_s \in \Lambda_s.
\end{equation}
Obviously, we have $d_{s,i}(\mathcal{L}^{(n,r_0)})=D(Q_{\mathcal{A}_i},X_i^n)$, where $Q_{\mathcal{A}_i}$ denotes the quantizer which maps each input sequence $x_i^n\in \mathcal{A}_i(\lambda_s)$ to $\lambda_s$, for $\lambda_s\in\Lambda_s$.
Since $\|x^n_i-\hat{x}_{c,i}^n\|\geq \|x^n_i-Q_c(x^n_i)\|$, it follows that
\begin{eqnarray*}
d_{c,i}(\mathcal{L}^{(n,r_0)})\geq  D(Q_c,X_i^n).
\end{eqnarray*}
Further, let us denote $\Delta_{i,sup}(\mathcal{L}^{(n,r_0)})\triangleq \sup_{x_i^n \in \mathbb{R}^n}\|x^n_i-\hat{x}_{c,i}^n\|$, $i=1,2$.
Additionally, let $\mathcal{P}_{e,SW}$ denote the probability that the Slepian-Wolf decoder fails.
In view of the definition  of $\mathcal{P}_{X_1X_2}({r_0})$ (see (\ref{pr0})) and  Proposition \ref{prop_dc}, it follows that, for  $i=1,2$,
\begin{eqnarray*}
d_{c,i}(\mathcal{L}^{(n,r_0)})\leq \frac{1}{n}(\mathcal{P}_{X_1X_2}({r_0}) + \mathcal{P}_{e,SW}) \Delta^2_{i,sup} + D(Q_c,X_i^n).
\end{eqnarray*}
The following lemma, proved in Appendix \ref{app:lemmas}, gives an upper bound for $\Delta_{i,sup}$.
\begin{lemma}
\label{d_sup}
There is some constant $\kappa_0$ such that, for each $i=1,2$, each positive integer $n$, and each LRDSC $\mathcal{L}^{(n,r_0)}$,
\begin{align*}
\Delta_{i,sup}(\mathcal{L}^{(n,r_0)}) \leq  \kappa_0\left(M\nu_s\right)^{\frac{1}{n}}.
\end{align*}
\end{lemma}
It is known that the probability that the Slepian-Wolf decoder fails can be made arbitrarily small by increasing the block length used for Slepian-Wolf encoding. Since $\Delta_{i,sup}(\mathcal{L}^{(n,r_0)})$ is bounded, it follows that the impact on the distortion of the Slepian-Wolf decoder failure can also be made arbitrarily small. Therefore, in the limit as the block length of the Slepian-Wolf encoder approaches infinity,
\begin{align}
D(Q_c,X_i^n) \leq d_{c,i}(\mathcal{L}^{(n,r_0)})&\leq \frac{1}{n}\kappa_0^2 \mathcal{P}_{X_1X_2}({r_0})\left(M\nu_s\right)^{\frac{2}{n}}\nonumber\\ & \quad + D(Q_c,X_i^n).\label{th3_eq01}
\end{align}
In order to evaluate the quantity $D(Q_c,X_i^n)$ at high resolution, we can directly use Lemma 1 in \cite{linder94}, which leads to
\begin{equation}
\label{centralQ}
D(Q_c,X_i^n)=G_c\nu_c^{\frac{2}{n}}(1+o(1)) \mbox{ as }\nu_c \rightarrow 0.
\end{equation}
Furthermore, in order to evaluate the rate, we need the following notation, for $i=1,2$,
\begin{equation*}
\mathcal{P}_i\triangleq \mathbb{P}[Q_{in}(Q_c(X_i^n))\in \mathcal{C}],
\end{equation*}
where $\mathcal{C}$ is defined in (\ref{calC}). We will use the following lemma, which is proved in Appendix \ref{app:lemmas}.
\begin{lemma}
\label{probGoTo0}
For $i=1,2$, we have $\lim\limits_{(\ref{lim0})}\mathcal{P}_i=0$.
\end{lemma}

\begin{IEEEproof}[Proof of Theorem \ref{th_main}]
Relation (\ref{dsi}) is proved in Appendix \ref{app:theorems}. Relation (\ref{th3_eq012}) follows based on (\ref{th3_eq01}) and (\ref{centralQ}). Let us prove now equality (\ref{R}). For this notice that the rate used to transmit $\beta_i(\lambda_i)$ is $\frac{1}{n}H(Q_{\mathcal{A}_i}({X}_i^n))$. The rate needed for $b_i$ is $\frac{1}{n}\left (-(1-\mathcal{P}_i)\log_2(1-\mathcal{P}_i) - \mathcal{P}_i \log_2 \mathcal{P}_i \right )$. The rate used for encoding $\tau_i$ equals  $\frac{1}{n}\mathcal{P}_i \log_2 |\mathcal{T}|= \mathcal{P}_i$. Finally, the rate needed for encoding $u_1$ and $u_2$ using Slepian-Wolf coding equals $\frac{1}{n}H(U_1,U_2)$.  As a consequence,
\begin{align}
\label{R0}
R(\mathcal{L}^{(n,r_0)})& =\frac{1}{n}\sum_{i=1}^2[H(Q_{\mathcal{A}_i}({X}_i^n))-(1-\mathcal{P}_i)\log_2(1-\mathcal{P}_i)\nonumber\\
& \quad + \mathcal{P}_i (-\log_2 \mathcal{P}_i +n)] + \frac{1}{n}H(U_1,U_2).
\end{align}
Since $\lim_{(\ref{lim0})}\bar{r}(\mathcal{A}_i({\bf{0}}))=0$, as shown in the proof of relation (\ref{dsi}),  we can apply Lemma \ref{csizar} stated at the end of this appendix, which is due to Csiszar \cite{csiszar73}. Thus, using the fact that $\nu(\mathcal{A}_i({\bf{0}}))=\nu_s$, we obtain that
\begin{equation} \label{R00}
\lim_{(\ref{lim0})}\frac{1}{n}\left (H(Q_{\mathcal{A}_i}({X}_i^n)) + \log_{2}\left(\nu_s\right)\right ) = h(X_i).
\end{equation}
Equations (\ref{R0}), (\ref{R00}) and Lemma \ref{probGoTo0} imply that
\begin{align*} 
&\lim_{(\ref{lim0})}\left (R(\mathcal{L}^{(n,r_0)}) + \frac{2}{n}\log_{2}\left(\nu_s\right) - \frac{1}{n}H(U_1,U_2)\right)\\
&= h(X_1) + h(X_2).
\end{align*}
Relation (\ref{R}) follows using the following equality, which is proved in Appendix \ref{app:theorems},
\begin{equation}
\label{lim_H(U)}
\lim_{(\ref{lim0})}H(U_i)=\log_2 K, \mbox{ for } i=1,2.
\end{equation}
Further, inequality (\ref{h(U2|U1)}) is based on $H(U_2|U_1)\leq H(U_2)=\log_2 K$, while inequality (\ref{cond_entropy}) is proved in Appendix \ref{app:theorems}. Finally, the claim that, in each of relations (\ref{dsi})-(\ref{R}) and (\ref{cond_entropy}),  the term hidden in the little-o notation  can be upperbounded by a function which does not depend on the joint pdf $f_{X_1X_2}$ and approaches $0$ under (\ref{lim0}) follows from the proofs of the aforementioned relations.
\end{IEEEproof}

\begin{IEEEproof}[Proof of Corrollary \ref{coro:new1}]
Notice that $(M\nu_s)^{\frac{2}{n}}=(M^2 K \nu_c)^{\frac{2}{n}}$. By plugging (\ref{pr0_cond1}) in
(\ref{th3_eq012}) and using the fact that $K$ and $\kappa_0$ are constants, relation  (\ref{final_dc1}) follows. Further, equalities  (\ref{dsi}) and (\ref{final_dc1}) imply that
\begin{align*}
d_{s,i}(\mathcal{L}^{(n,r_0)})d_{c,i}(\mathcal{L}^{(n,r_0)})  & = \frac{1}{4}G_{s/2}G_c(M\nu_s\nu_c)^{\frac{2}{n}}(1+o(1))\\ & =
\frac{1}{4}G_{s/2}G_c\left( \frac{\nu_s^2}{K}\right)^{\frac{2}{n}}(1+o(1)).
\end{align*}
By substituting this in (\ref{R}), relation (\ref{rate_coro_new1}) follows.

In order to prove (\ref{rate_coro_new2}), we first apply Fano's inequality and obtain that
\begin{equation}
\label{h_e}
H(U_2|U_1) \leq H_b(\mathbb{P}[U_1\neq U_2]) + \mathbb{P}[U_1\neq U_2] \log_2 K,
\end{equation}
where $H_b(\cdot)$ denotes the binary entropy function.
Next we assume that $r_0\leq r_c$ and use the following inequality proved in Appendix \ref{app:theorems} (in the proof of relation (\ref{cond_entropy}))
\begin{equation*}
\mathbb{P}[U_1\neq U_2] \leq  1-\left(1-\frac{r_0}{r_c}\right)^n + \mathcal{P}_{X_1X_2}(r_0) + o(1),
\end{equation*}
where the term hidden in the little-o notation does not depend on the joint pdf $f_{X_1X_2}$. The fact that $\lim\limits_{(\ref{lim0})}\frac{r_0}{r_c}=0$, together with $\lim\limits_{(\ref{lim0})}\mathcal{P}_{X_1X_2}(r_0)=0$, further implies that
$\lim\limits_{(\ref{lim0})}\mathbb{P}[U_1\neq U_2]=0$. Combining this with (\ref{h_e}) leads to  $\lim\limits_{(\ref{lim0})}H(U_2|U_1)=0$.
By applying this result in (\ref{rate_coro_new1}), relation
(\ref{rate_coro_new2}) follows.
\end{IEEEproof}

\begin{lemma}[Csiszar \cite{csiszar73}]
\label{csizar}
\footnote{The statement of this lemma is taken from \cite{linder94}.}
Let $Z=(Z_1,\cdots,Z_k)$ be an $\mathbb{R}^k$ valued random vector with density $f_Z$. Suppose that there exists some Borel measurable partition $\mathcal{B}_0=\{B_1,B_2,\cdots \}$ of $\mathbb{R}^k$ into sets of finite Lesbesgue measure such that
\begin{equation*}
-\sum_n \mathbb{P}[Z\in B_n]\log \mathbb{P}[Z\in B_n] <\infty.
\end{equation*}
Suppose furthermore, that for some $\rho>0$, some positive integer $s$, and for all $k$, the distance of $B_k$ from any other $B_l$ is greater than $\rho$ for all but at most $s$ indexes $l$. Let $\mathcal{A}=\{A_0,A_1,\cdots\}$ be a measurable partition with equal Lesbegue measure, i.e., $\lambda(A_i)=\epsilon$, $i=1,2,\cdots$, and let us denote the supremum of the diameters of the sets $A_i$ by $\delta(\mathcal{A})$. Then we have
\begin{equation*}
\lim\limits_{\delta(\mathcal{A})\rightarrow 0}\left( H_{\mathcal{A}}(Z) +\log \epsilon \right) = h(f_Z),
\end{equation*}
where
\begin{equation*}
H_{\mathcal{A}}(Z) = -\sum_n\mathbb{P}[Z\in A_n] \log \mathbb{P}[Z\in A_n],
\end{equation*}
and
\begin{equation*}
h(f_Z) = -\int_{\mathbb{R}^k}f_Z(x^k)\log f_Z(x^k) \ dx^k,
\end{equation*}
the differential entropy of $Z$. Moreover, if $Z$ has no density, then the above limit is $-\infty$. It should be mentioned that with the above conditions $h(f_Z)$ is always well-defined and $h(f_Z)<\infty$.
\end{lemma}

\section{Proof of Relations (\ref{dsi}), (\ref{lim_H(U)}) and (\ref{cond_entropy})}
\label{app:theorems}

\begin{IEEEproof}[Proof of Relation (\ref{dsi})] First let us fix $i$. We will split the proof into two parts. In Part 1 we show that if  $\lim\limits_{(\ref{lim0})}\frac{G(\mathcal{A}_i({\bf{0}}))}{M^{\frac{2}{n}}}$ exists, then
\begin{equation}
\label{part1}
\lim_{(\ref{lim0})}\frac{D(Q_{\mathcal{A}_i},X_i^n)}{(M\nu_s)^{\frac{2}{n}}}=
\lim_{(\ref{lim0})}\frac{G(\mathcal{A}_i({\bf{0}}))}{M^{\frac{2}{n}}}.
\end{equation}
In Part 2 we prove that
\begin{equation}
\label{part2}
\lim_{(\ref{lim0})}\frac{G(\mathcal{A}_i({\bf{0}}))}{M^{\frac{2}{n}}}=
\frac{1}{4}G_{s/2}.
\end{equation}
\noindent {\bf Part 1.}\footnote{This proof uses ideas from the proof of \cite[Lemma 1]{linder94}.} The proof is based on the idea that, in the limit of (\ref{lim0}), the pdf $f_{X_i^n}$ can be approximated by a uniform density function over each  set $\mathcal{A}_i(\lambda_s)$.
This density function is $f_{\theta,\mu}:\mathbb{R}^n\rightarrow [0,\infty)$ defined as follows. For each $\lambda_s\in \Lambda_s$ and $x^n\in \mathcal{A}_i(\lambda_s)$, let
\begin{align*}
f_{\theta,\mu}(x^n)&=  \frac{\mathbb{P}[X_i^n\in \mathcal{A}_i(\lambda_s)]}{\nu(\mathcal{A}_i(\lambda_s))} \\
&= \frac{1}{\nu(\mathcal{A}_i(\lambda_s))}\int_{\mathcal{A}_i(\lambda_s)} f_{X_i^n}(y^n)dy^n.
\end{align*}
Let $X_{\theta,\mu}^n$ denote the random variable with pdf $f_{\theta,\mu}$.
Note that
\begin{align}
\label{diff_dist}
&| D(Q_{\mathcal{A}_i},X_{\theta,\mu}^n)-D(Q_{\mathcal{A}_i},X_i^n)| \nonumber \\ & \leq \frac{1}{n}\sum_{\lambda_s\in \Lambda_s}\int_{\mathcal{A}_i(\lambda_s)}\|x^n-\lambda_s \|^2 |f_{\theta,\mu}(x^n)-f_{X_i^n}(x^n)|dx^n \nonumber \\
& \leq\frac{1}{n}\sum_{\lambda_s\in \Lambda_s}\bar{r}(\mathcal{A}_i({\bf{0}}))^2\int_{\mathcal{A}_i(\lambda_s)}|f_{\theta,\mu}(x^n)-f_{X_i^n}(x^n)|dx^n  \nonumber \\
&=\frac{\bar{r}(\mathcal{A}_i({\bf{0}}))^2}{n}\int_{\mathbb{R}^n} |f_{\theta,\mu}(x^n)-f_{X_i^n}(x^n)|dx^n,
\end{align}
where the second inequality is based on the fact that $\mathcal{A}_i(\lambda_s)=\lambda_s + \mathcal{A}_i({\bf{0}})$, which implies that $\max\limits_{x^n \in \mathcal{A}_i(\lambda_s)} \|x^n-\lambda_s \|^2 = \bar{r}(\mathcal{A}_i({\bf{0}}))$.
Let us analyze now the quantity  $\bar{r}(\mathcal{A}_i({\bf{0}}))$.
Recall that $\mathcal{A}_i({\bf{0}})= \cup_{\lambda \in \beta_{i}^{-1}({\bf{0}})} (\lambda +\mathcal{M}({\bf{0}}))$, where $\mathcal{M}({\bf{0}})\triangleq \cup_{\lambda_c \in V_{in}({\bf{0}})\cap \Lambda_c} V_c(\lambda_c)$. Then it follows that
\begin{equation}
\label{r_sc}
\bar{r}(\mathcal{A}_i({\bf{0}})) \leq \bar{r}(\beta_i^{-1}({\bf{0}})) + \bar{r}(\mathcal{M}({\bf{0}})).
\end{equation}
Further,
\begin{equation}
\label{bar_r_M0}
\bar{r}(\mathcal{M}({\bf{0}})) \leq \bar{r}_{in}+ \bar{r}_c \leq 2\bar{r}_{in} = 2\theta\bar{r}_{in,0}.
\end{equation}
Since we are interested in computing the limits in (\ref{part1}) under (\ref{lim0}), we may assume that $\mu$ is conveniently large. In particular, in the sequel we will assume that $\mu\geq 8$ so that relation (\ref{bar_r}) leads to
\begin{equation}
\label{bar_r_beta1}
\bar{r}(\beta_i^{-1}({\bf{0}})) \leq \mu^2\theta\bar{r}_{in,0}.
\end{equation}
Finally, relations (\ref{r_sc})-(\ref{bar_r_beta1}), together with the fact that  $M=\mu^n$ and $\nu_s = \mu^n\theta^n \nu_{in,0}$, lead to
\begin{equation}
\label{r_sc1}
\frac{\bar{r}(\mathcal{A}_i({\bf{0}}))}{(M\nu_s)^{\frac{1}{n}}} \leq \frac{2\theta\bar{r}_{in,0} +\mu^2\theta\bar{r}_{in,0}}{\mu^2\theta \nu_{in,0}^{\frac{1}{n}}}\rightarrow
\frac{\bar{r}_{in,0}}{2\nu_{in,0}^{\frac{1}{n}}}
\end{equation}
in the limit of (\ref{lim0}). The above result also implies that $\bar{r}(\mathcal{A}_i({\bf{0}}))\rightarrow 0$ under (\ref{lim0}). This enables us to apply Lemma \ref{lemma1}, which is stated and proved in Appendix \ref{app:lemmas}, and we obtain that $f_{\theta,\mu}(x^n)\rightarrow f_{X_1}^n(x^n)$, $x^n \in \mathbb{R}^n$, under (\ref{lim0}). Using further Scheffe's theorem \cite{scheffe}, it follows that $\int_{\mathbb{R}^n} |f_{\theta,\mu}(x^n)-f_{X_1^n}(x^n)|dx^n \rightarrow 0$  under (\ref{lim0}). Combining this further with (\ref{diff_dist}) and (\ref{r_sc1}) gives
\begin{eqnarray}
\label{diff_dist1}
&\lim\limits_{(\ref{lim0})}\frac{1}{(M\nu_s)^{\frac{2}{n}}}| D(Q_{\mathcal{A}_i},X_{\theta,\mu}^n)-D(Q_{\mathcal{A}_i},X_i^n)|=0.
\end{eqnarray}
Using now the fact that  $f_{\theta,\mu}$ is uniform over each quantizer cell $\mathcal{A}_i(\lambda_s)$, we obtain that
\begin{align}
\nonumber
& D(Q_{\mathcal{A}_i},X_{\theta,\mu}^n)=\frac{1}{n}\sum_{\lambda_s\in \Lambda_s}\int_{\mathcal{A}_i(\lambda_s)}\|x^n-\lambda_s\|^2f_{\theta,\mu}(x^n)dx^n \\\nonumber
&\qquad =\frac{1}{n}\sum_{\lambda_s\in \Lambda_s}\frac{\mathbb{P}[X_i^n\in \mathcal{A}_i(\lambda_s)]}{\nu(\mathcal{A}_i(\lambda_s))}\int_{\mathcal{A}_i(\lambda_s)}\|x^n-\lambda_s\|^2dx^n  \\\nonumber
& \qquad \stackrel{(a)}{=} \frac{1}{n\nu(\mathcal{A}_i({\bf{0}}))}\int_{\mathcal{A}_i({\bf{0}})}\|x^n\|^2dx^n \sum_{\lambda_s\in \Lambda_s}\mathbb{P}[X_i^n\in \mathcal{A}_i(\lambda_s)] \\\nonumber
& \qquad = \frac{1}{n\nu(\mathcal{A}_i({\bf{0}}))}\int_{\mathcal{A}_i({\bf{0}})}\|x^n\|^2dx^n \mathbb{P}[X_i^n\in \mathbb{R}^n] \\ \nonumber
& \qquad = G(\mathcal{A}_i({\bf{0}}))(\nu(\mathcal{A}_i({\bf{0}})))^{\frac{2}{n}}\\
& \qquad\stackrel{(b)}{=}G(\mathcal{A}_i({\bf{0}}))\nu_s^{\frac{2}{n}},
\label{part1_finaleq}
\end{align}
where (a) uses the fact that $\mathcal{A}_i(\lambda_s)=\lambda_s + \mathcal{A}_i({\bf{0}})$,  while (b) is based on the fact that $\nu(\mathcal{A}_i({\bf{0}}))=\nu_s$ since $\mathcal{A}_i({\bf{0}})$ is a fundamental cell of the lattice $\Lambda_s$. Relations (\ref{diff_dist1}) and (\ref{part1_finaleq}) prove the claim of Part 1.\\
\noindent {\bf Part 2.}
In order to prove (\ref{part2}),
we will first evaluate $\int_{\mathcal{A}_i({\bf{0}})}\| x^n\|^{2}dx^n$.
Using the fact that $\mathcal{A}_i({\bf{0}})= \cup_{\lambda \in \beta_{i}^{-1}({\bf{0}})} (\lambda +\mathcal{M}({\bf{0}}))$ and relation (\ref{shiftbeta-1of0}), we  obtain that
\begin{equation}
\label{ai0}
\mathcal{A}_i({\bf{0}}) = \cup_{\lambda \in \mathcal{U}}\left ( \lambda - \beta_i(\lambda) + \mathcal{M}({\bf{0}}) \right).
\end{equation}
Using further Lemma \ref{lemma:shift}, which is stated and proved in Appendix \ref{app:lemmas}, we obtain that
\begin{align}
& \int_{\lambda - \beta_i(\lambda) + \mathcal{M}({\bf{0}})} \|x^n\|^2 dx^n = \|\lambda - \beta_i(\lambda)\|^2 \nu(\mathcal{M}({\bf{0}}))
 \nonumber \\
& \quad + 2\langle \int_{\mathcal{M}({\bf{0}})}x^ndx^n,\lambda - \beta_i(\lambda)\rangle + \int_{\mathcal{M}({\bf{0}})}\|x^n\|^2 dx^n. \label{ai01}
\end{align}
It is easy to see that $\mathcal{M}({\bf{0}})$ is a fundamental cell of the lattice $\Lambda_{in}$, therefore, $\nu(\mathcal{M}({\bf{0}}))=\nu_{in}$. Further, relations (\ref{ai0}) and (\ref{ai01}) lead to
\begin{align*}
\int_{\mathcal{A}_i({\bf{0}})} \|x^n\|^2 dx^n & =
\underbrace{|\mathcal{U}|\int_{\mathcal{M}({\bf{0}})}\|x^n\|^2 dx^n}_{T_1} \\
& \quad  + \underbrace{2\sum_{\lambda \in \mathcal{U}}\langle \int_{\mathcal{M}({\bf{0}})}x^ndx^n, \lambda - \beta_i(\lambda)\rangle}_{T_{2,i}} \\
& \quad + \underbrace{\nu_{in}\sum_{\lambda \in \mathcal{U}} \|\lambda - \beta_i(\lambda)\|^2}_{T_{3,i}}.
\end{align*}
Then
\begin{align}
\frac{G(\mathcal{A}_i({\bf{0}}))}{M^{\frac{2}{n}}}&=\frac{T_1}{nM^{\frac{2}{n}}(M\nu_{in})^{1+\frac{2}{n}}} + \frac{T_{2,i}}{nM^{\frac{2}{n}}(M\nu_{in})^{1+\frac{2}{n}}} \nonumber \\ & \quad + \frac{T_{3,i}}{nM^{\frac{2}{n}}(M\nu_{in})^{1+\frac{2}{n}}}.
\label{eq:222}
\end{align}
We will prove first that the first two terms on the right hand side of the above equality approach $0$ in the limit of (\ref{lim0}).
Consider the first term. Note that $\int_{\mathcal{M}({\bf{0}})}\|x^n\|^2 dx^n\leq \left(\bar{r}(\mathcal{M}({\bf{0}}))\right)^2 \nu_{in}$.
Combining this further with (\ref{bar_r_M0}) and the fact that $|\mathcal{U}|=M$ gives
\begin{align}
\frac{T_1}{nM^{\frac{2}{n}}(M\nu_{in})^{1+\frac{2}{n}}}&\leq
\frac{4M\theta^2 \bar{r}^2_{in,0} \nu_{in}}{nM^{\frac{2}{n}}(M\nu_{in})^{1+\frac{2}{n}}}\nonumber \\
& = \frac{4\bar{r}^2_{in,0}}{nM^{\frac{4}{n}}\nu_{in,0}^{\frac{2}{n}}}\rightarrow 0 \mbox{ under (\ref{lim0}).}
\label{T1}
\end{align}
It is easy to see that the closure of a lattice Voronoi cell of the origin is symmetric about the origin. Therefore, if $\Lambda_{in}$ is a clean sublattice of $\Lambda_c$, i.e., there are no points of $\Lambda_c$ on the boundary of $V_{in}({\bf{0}})$, then the set $\Lambda_c\cap V_{in}({\bf{0}})$ is symmetric about the origin. The above considerations further imply that the closure of the set $\mathcal{M}({\bf{0}})$ is symmetric about the origin, thus $\int_{\mathcal{M}({\bf{0}})}x^ndx^n=0$. Then the second term in (\ref{eq:222}) is $0$. When $\Lambda_{in}$ is not a clean sublattice of $\Lambda_c$, the aforementioned term still approaches $0$ in the limit of (\ref{lim0}), as we prove next. Note that
\begin{align}
|T_{2,i}|&= 2\left | \sum_{\lambda \in \mathcal{U}} \int_{\mathcal{M}({\bf{0}})} \langle x^n, \lambda - \beta_i(\lambda)\rangle dx^n\right |\nonumber\\
&\leq  2 \sum_{\lambda \in \mathcal{U}} \int_{\mathcal{M}({\bf{0}})} \left |\langle x^n, \lambda - \beta_i(\lambda)\rangle \right | dx^n\nonumber\\
&\stackrel{(a)}{\leq}  2 \sum_{\lambda \in \mathcal{U}} \int_{\mathcal{M}({\bf{0}})} \left \|x^n \right \|
\left \| \lambda-\beta_i(\lambda) \right \|  dx^n \nonumber\\
& =   2  \int_{\mathcal{M}({\bf{0}})}\left \|x^n\right \| dx^n  \sum_{\lambda \in \mathcal{U}}
\left \| \lambda-\beta_i(\lambda) \right \| \nonumber\\
&\stackrel{(b)}{\leq}  2 \bar{r}(\mathcal{M}({\bf{0}}))\nu_{in}M (\max_{\lambda \in \mathcal{U}}\|\lambda\| +
\max_{\lambda \in \mathcal{U}}\|\beta_i(\lambda)\|) \nonumber\\
&\stackrel{(c)}{\leq}  4\theta\bar{r}_{in,0}\nu_{in}M\mu^2\theta\bar{r}_{in,0}  \stackrel{(d)}{=}  4 \theta^2\nu_{in}M^{1+\frac{2}{n}}\bar{r}_{in,0}^2.\label{t2i}
\end{align}
Here (a) follows from the Cauchy-Schwarz inequality and (b) is based on the fact that
$\int_{\mathcal{M}({\bf{0}})}\|x^n\|dx^n \leq \bar{r}(\mathcal{M}({\bf{0}})) \nu_{in}$ and $|\mathcal{U}|=M$; additionally, (c) follows from (\ref{bar_r_M0}) and the discussion in the paragraph below equation (\ref{bar_r_beta}); finally, (d) is based on the fact that $\mu=M^{\frac{1}{n}}$. Further, relation (\ref{t2i}) implies that
\begin{align}
\frac{|T_{2,i}|}{nM^{\frac{2}{n}}(M\nu_{in})^{1+\frac{2}{n}}} & \leq  \frac{4 \theta^2\nu_{in}M^{1+\frac{2}{n}}\bar{r}_{in,0}^2}{nM^{1+\frac{4}{n}}\nu_{in}\theta^2 \nu_{in,0}^{\frac{2}{n}}} \nonumber \\
& = \frac{4 \bar{r}_{in,0}^2}{nM^{\frac{2}{n}} \nu_{in,0}^{\frac{2}{n}}}
\rightarrow 0 \mbox{ under (\ref{lim0}).}
\label{t2i2}
\end{align}
Let us evaluate now $\frac{T_{3,i}}{\nu_{in}}$.
We need to treat separately the cases $i=1$ and $i=2$. Recall that  $\mathcal{U}= \cup_{\tau \in \mathcal{T}}V_{s/2}(\tau)\cap \Lambda_{in}$. We will denote $\hat{V}_{s/2}(\tau)\triangleq V_{s/2}(\tau)\cap \Lambda_{in}$. Using further (\ref{def_bi_U}), we obtain that
\begin{align}
\frac{T_{3,1}}{\nu_{in}}
& =  \sum_{\tau \in \mathcal{T}}\sum_{\lambda \in \hat{V}_{s/2}(\tau )}\| \lambda - \mu(\lambda-\tau)\|^{2} \nonumber\\
& =  \sum_{\tau \in \mathcal{T}}\sum_{\lambda \in \hat{V}_{s/2}(\tau )}\| (1-\mu)(\lambda-\tau) + \tau \|^{2} \nonumber\\
& =  \sum_{\tau \in \mathcal{T}}\sum_{\lambda \in \hat{V}_{s/2}(\tau )} ( \|(1-\mu)(\lambda-\tau)\|^{2} + \| \tau \|^{2} \nonumber\\
& \qquad \qquad \qquad \quad +  2 \langle (1-\mu)(\lambda-\tau),\tau\rangle  ) \nonumber\\
& =  \sum_{\tau \in \mathcal{T}}\sum_{\lambda \in \hat{V}_{s/2}(\tau )} (1-\mu)^2\| \lambda-\tau\|^{2} + \sum_{\tau \in \mathcal{T}}\sum_{\lambda \in \hat{V}_{s/2}(\tau )} \| \tau \|^{2} \nonumber\\
& \quad  +  \underbrace{2\sum_{\tau \in \mathcal{T}}\sum_{\lambda \in \hat{V}_{s/2}(\tau )}\langle (1-\mu)(\lambda-\tau),\tau\rangle}_{T_4} \nonumber\\
& \stackrel{(a)}{=}  \underbrace{(1-\mu)^2|\mathcal{T}|\sum_{\lambda \in \hat{V}_{s/2}({\bf{0}})} \| \lambda\|^{2}}_{T_5} + \underbrace{\frac{M}{|\mathcal{T}|}\sum_{\tau \in \mathcal{T}}\| \tau \|^{2}}_{T_6} + T_4,
\label{t31}
\end{align}
where (a) is based on the fact that $\hat{V}_{s/2}(\tau )=\tau + \hat{V}_{s/2}({\bf{0}})$ and $|\hat{V}_{s/2}({\bf{0}})|=\frac{M}{|\mathcal{T}|}$.
Relation (\ref{t31}) leads to
\begin{align}
&\frac{T_{3,1}}{nM^{\frac{2}{n}}(M\nu_{in})^{1+\frac{2}{n}}} \nonumber\\
&=
\frac{T_{4}}{nM^{1+\frac{4}{n}}\nu_{in}^{\frac{2}{n}}} + \frac{T_{5}}{nM^{1+\frac{4}{n}}\nu_{in}^{\frac{2}{n}}} + \frac{T_{6}}{nM^{1+\frac{4}{n}}\nu_{in}^{\frac{2}{n}}}.\label{t311}
\end{align}
We will show first that the first and last terms on the right hand side of (\ref{t311}) approach $0$ in the limit of (\ref{lim0}). For this we need to introduce the following notation.
For any two nested lattices $\Lambda_2\subset \Lambda_1$  in $\mathbb{R}^n$, denote $\mathcal{C}_{\Lambda_2:\Lambda_1}\triangleq
\cup_{\lambda_1\in V_{\Lambda_2}({\bf{0}})\cap \Lambda_1}V_{\Lambda_1} (\lambda_1)$.
Using  Lemma \ref{sum_squared_norms}, which is stated and proved in Appendix \ref{app:lemmas},  we obtain
\begin{align*}
\frac{T_6}{nM^{1+\frac{4}{n}}\nu_{in}^{\frac{2}{n}}}&=
\frac{\frac{M}{2^n}n2^n}{nM^{1+\frac{4}{n}}\nu_{in}^{\frac{2}{n}}}\left(G(\mathcal{C}_{\Lambda_s:\Lambda_{s/2}})\nu_{s}^{\frac{2}{n}}- G_{s/2}\nu_{{s/2}}^{\frac{2}{n}}\right)\nonumber \\
&= \frac{1}{M^{\frac{2}{n}}}\left(G(\mathcal{C}_{\Lambda_s:\Lambda_{s/2}})- \frac{1}{4}G_{s/2}\right),
\end{align*}
where the last equality is based on $\nu_s=M\nu_{in}$ and  $\nu_{s/2}=M\nu_{in}/2^n$. As the parameters $\mu$ and $\theta$ vary, both lattices $\Lambda_s$ and $\Lambda_{s/2}$ are scaled by the same factor, therefore the set $\mathcal{C}_{\Lambda_s:\Lambda_{s/2}}$ is scaled by that factor. Since the second moment is invariant under scaling, it follows that $G(\mathcal{C}_{\Lambda_s:\Lambda_{s/2}})- \frac{1}{4}G_{s/2}$ remains constant as $\theta$ and $\mu$ vary.
Consequently,
\begin{eqnarray}
\lim_{(\ref{lim0})}\frac{T_6}{nM^{1+\frac{4}{n}}\nu_{in}^{\frac{2}{n}}}=
0.
\end{eqnarray}
Consider now the first term on the right hand side of (\ref{t311}). We have
\begin{align*}
|T_4|&\leq   2|\mu-1|\sum_{\tau \in \mathcal{T}}\sum_{\lambda \in \hat{V}_{s/2}(\tau )}|\langle \lambda-\tau,\tau\rangle|\\
&\stackrel{(a)}{\leq} 2|\mu-1|\sum_{\tau \in \mathcal{T}}\sum_{\lambda \in \hat{V}_{s/2}(\tau )}\|\lambda-\tau\| \|\tau \|\\
& \leq  2|\mu-1|M \max_{\tau \in\mathcal{T}}\|\tau \|\max_{\lambda \in \hat{V}_{s/2}({\bf{0}})}\|\lambda\|\\
& \leq  2\mu M \bar{r}_s\bar{r}_{s/2},
\end{align*}
where (a) is based on the Cauchy-Schwartz inequality. Using further the fact that $\mu=M^{\frac{1}{n}}$, while $\bar{r}_{s/2}=\bar{r}_{s}/2=M^{\frac{1}{n}}\bar{r}_{in}/2$, leads to
\begin{align}
\frac{|T_4|}{nM^{1+\frac{4}{n}}\nu_{in}^{\frac{2}{n}}}&\leq
\frac{M^{1+\frac{3}{n}}\bar{r}_{in}^2}{nM^{1+\frac{4}{n}}\nu_{in}^{\frac{2}{n}}}\nonumber\\
 &= \frac{\bar{r}_{in,0}^2}{nM^{\frac{1}{n}}\nu_{in,0}^{\frac{2}{n}}}
\rightarrow 0 \mbox{ as (\ref{lim0})  holds}.\label{t4}
\end{align}
In order to evaluate the second term in (\ref{t311}), we use again Lemma
\ref{sum_squared_norms} and obtain that
\begin{align*}
\frac{T_5}{nM^{1+\frac{4}{n}}\nu_{in}^{\frac{2}{n}}}&=
\frac{(\mu-1)^2nM}{nM^{1+\frac{4}{n}}\nu_{in}^{\frac{2}{n}}}\left(G(\mathcal{C}_{\Lambda_{s/2}:\Lambda_{in}})\nu_{s/2}^{\frac{2}{n}}
- G_{in}\nu_{in}^{\frac{2}{n}}\right)\nonumber \\
&= \frac{(M^{\frac{1}{n}}-1)^2}{M^{\frac{4}{n}}}\left(G(\mathcal{C}_{\Lambda_{s/2}:\Lambda_{in}})\frac{M^{\frac{2}{n}}}{4}
- G_{in}\right),
\end{align*}
where the last equality relies on the fact that $\mu=M^{\frac{1}{n}}$, while $\nu_{s/2}=M\nu_{in}/2^n$. Further, we obtain that
\begin{eqnarray}
\label{t5final}
\lim_{(\ref{lim0})}\frac{T_5}{nM^{1+\frac{4}{n}}\nu_{in}^{\frac{2}{n}}}=
\lim_{(\ref{lim0})}\frac{G(\mathcal{C}_{\Lambda_{s/2}:\Lambda_{in}})}{4}=\frac{G_{s/2}}{4},
\end{eqnarray}
where the last equality follows from Lemma \ref{shape}, which is stated and proved in Appendix \ref{app:lemmas}.

Relations (\ref{t311})-(\ref{t5final}) imply that
\begin{eqnarray}
\label{lim2}
\lim_{(\ref{lim0})}\frac{T_{3,1}}{nM^{1+\frac{4}{n}}\nu_{in}^{\frac{2}{n}+1}}=\frac{1}{4} G_{s/2}.
\end{eqnarray}
Combining the above with (\ref{eq:222}), (\ref{T1}) and (\ref{t2i2}), we obtain that (\ref{part2}) holds for $i=1$.
In order to prove the claim for $i=2$,
we need to evaluate now $\frac{T_{3,2}}{\nu_{in}}$. Note that
\begin{align*}
\frac{T_{3,2}}{\nu_{in}}
& =  \sum_{\tau \in \mathcal{T}}\sum_{\lambda \in \hat{V}_{s/2}(\tau )}\| \lambda - 2\tau + \mu(\lambda-\tau)\|^{2}\\
& =  \sum_{\tau \in \mathcal{T}}\sum_{\lambda \in \hat{V}_{s/2}(\tau )}\| (1+\mu)(\lambda-\tau) - \tau \|^{2}\\
& =  \sum_{\tau \in \mathcal{T}}\sum_{\lambda \in \hat{V}_{s/2}(\tau )} ( \| (1+\mu)(\lambda-\tau)\|^{2} + \| \tau \|^{2} \\
& \qquad  \qquad  \qquad \quad + 2 \langle (1+\mu)(\lambda-\tau),\tau\rangle ) \\
& = \sum_{\tau \in \mathcal{T}}\sum_{\lambda \in \hat{V}_{s/2}(\tau )}\left ( \| (1+\mu)(\lambda-\tau)\|^{2} + \| \tau \|^{2}  \right )\\
& \quad + 2\sum_{\tau \in \mathcal{T}}\sum_{\lambda \in \hat{V}_{s/2}(\tau )}\langle (1+\mu)(\lambda-\tau),\tau\rangle.
\end{align*}
Next the conclusion follows using similar arguments as for $i=1$. This observation concludes the proof.
\end{IEEEproof}

\begin{IEEEproof}[Proof of Relation (\ref{lim_H(U)})]
In order to prove the claim, we will show that $U_i$ approaches a uniform distribution. To prove this let $u\in V_{in}({\bf{0}})\cap \Lambda_{c}$.

The general idea of the proof is that, as the limits of (\ref{lim0}) are approached,  the pdf $f_{X_i^n}$ can be approximated by a pdf which is uniform  on each set $\mathcal{M}(\lambda)$.
Then in the limit of (\ref{lim0}),
\begin{align*}
\mathbb{P}[U_i=u]&=\sum_{\lambda \in \Lambda_{in}}\int_{V_c(\lambda + u)}f_{X_i^n}(x^n) dx^n \\
& \stackrel{(a)}{\approx}	\sum_{\lambda \in \Lambda_{in}} f_{X_i^n}(\lambda)\nu_c \\
&=  \sum_{\lambda \in \Lambda_{in}} f_{X_i^n}(\lambda)\frac{\nu_{in}}{K} \\
& \stackrel{(b)}{\approx} \frac{1}{K} \sum_{\lambda \in \Lambda_{in}}\int_{\mathcal{M}(\lambda)}f_{X_i^n}(x^n) dx^n \\
& = \frac{1}{K}.
\end{align*}
Next we provide a rigorous treatment of relations (a) and (b).

Define a density function $f_{\theta,\mu}:\mathbb{R}^n\rightarrow [0,\infty)$, which is uniform on each set $\mathcal{M}(\lambda)$, as follows
\begin{equation*}
f_{\theta,\mu}(x^n)=\frac{1}{\nu(\mathcal{M}(\lambda))}\int_{\mathcal{M}(\lambda)} f_{X_i^n}(y^n)dy^n,\quad x^n\in \mathcal{M}(\lambda).
\end{equation*}
Then in view of Lemma \ref{lemma1} (stated and proved in Appendix \ref{app:lemmas}), we have that $f_{\theta,\mu}(x^n)\rightarrow f_{X_i^n}(x^n)$, $x^n\in \mathbb{R}^n$, under (\ref{lim0}). Further, we have
\begin{align}
\mathbb{P}[U_i=u]&=\int_{\cup_{\lambda \in \Lambda_{in}}V_c(\lambda + u)}\left(f_{X_i^n}(x^n)-f_{\theta,\mu}(x^n)\right)dx^n\nonumber \\ & \quad +\int_{\cup_{\lambda \in \Lambda_{in}}V_c(\lambda + u)}f_{\theta,\mu}(x^n)dx^n\nonumber \\
&\leq   \int_{\cup_{\lambda \in \Lambda_{in}}V_c(\lambda + u)}|f_{X_i^n}(x^n)-f_{\theta,\mu}(x^n)|dx^n \nonumber \\
&\quad+
\sum_{\lambda \in \Lambda_{in}}\int_{V_c(\lambda + u)}f_{\theta,\mu}(x^n)dx^n.
\label{coro2E2}
\end{align}
Note that
\begin{align*}
& \int_{\cup_{\lambda \in \Lambda_{in}}V_c(\lambda + u)}|f_{X_i^n}(x^n)-f_{\theta,\mu}(x^n)|dx^n  \\
& \qquad \leq
\int_{\mathbb{R}^n}|f_{X_i^n}(x^n)-f_{\theta,\mu}(x^n)|dx^n \rightarrow 0 \mbox{ under (\ref{lim0})},
\end{align*}
where the last relation is valid in view of Scheffe's theorem \cite{scheffe}.

Further, since $f_{\theta,\mu}$ is constant on each $\mathcal{M}(\lambda)$, we have
\begin{align}
\sum_{\lambda \in \Lambda_{in}}\int_{V_c(\lambda + u)}f_{\theta,\mu}(x^n)dx^n & =
\sum_{\lambda \in \Lambda_{in}} f_{\theta,\mu}(\lambda)\frac{\nu_{in}}{K} \nonumber \\
& = \frac{1}{K} \sum_{\lambda \in \Lambda_{in}}\int_{\mathcal{M}(\lambda)}f_{X_i^n}(x^n) dx^n \nonumber \\&= \frac{1}{K}.
\label{coro2E1}
\end{align}
Relations (\ref{coro2E2})-(\ref{coro2E1}), together with the fact that the size of the alphabet of $U_i$ is $K$ and $K$ is constant, prove the claim. With this observation the proof is complete.
\end{IEEEproof}

\begin{IEEEproof}[Proof of Relation (\ref{cond_entropy})]
Using a variant of Fano's inequality, we obtain that
\begin{align}
H(U_2|U_1)
& \leq  1 + \mathbb{P}[U_1\neq U_2]\log_2 K,\label{eq:e00}
\end{align}
where we used  the fact that $H(U_2)=\log_2 K$.
Let $\lambda_{c,1}=Q_c(x_1^n)$.
Notice that if $x^n_2-x_1^n\in\mathcal{B}(r_0)$ and the distance from $x_1^n$ to the boundary of the Voronoi cell $V_c(\lambda_{c,1})$ is larger than or equal to $r_0$, then it is guaranteed that $x_2^n\in V_c(\lambda_{c,1})$, thus $u_2=u_1$. Now let us denote
\begin{align*}
\mathcal{E}(\lambda_c)\triangleq V_c(\lambda_c) \setminus \left(1 - \frac{r_0}{r_c}\right)V_c(\lambda_c),
\end{align*}
for each $\lambda_c\in \Lambda_c$, and $\mathcal{E}\triangleq \cup_{\lambda_c \in \Lambda_c}\mathcal{E}(\lambda_c)$.
It follows that
\begin{equation}
\label{eq:e0}
\mathbb{P}[U_1\neq U_2] \leq \mathcal{P}_{X_1X_2}(r_0) + \mathbb{P}[X_1^n \in \mathcal{E}].
\end{equation}
Further, we obtain
\begin{align}
\label{eq:e1}
\mathbb{P}[X_1^n \in \mathcal{E}] \leq  \int_{\mathcal{E}} |f_{X_1^n}(x^n)-f_{\theta,c}(x^n)|dx^n + \int_{\mathcal{E}} f_{\theta,c}(x^n) dx^n,
\end{align}
where $f_{\theta,\mu}$ was defined in the proof of relation (\ref{lim_H(U)}). According to that proof, the first integral in (\ref{eq:e1}) approaches $0$ in the limit of (\ref{lim0}). Since $f_{\theta,\mu}$ is uniform over each Voronoi region of the central lattice, we have
\begin{align}
\int_{\mathcal{E}} f_{\theta,\mu}(x^n) dx^n & =
\sum_{\lambda_c \in \Lambda_c}\int_{ \mathcal{E}(\lambda_c)}f_{\theta,\mu}(x^n) dx^n \nonumber\\
& = \sum_{\lambda_c \in \Lambda_c}f_{\theta,\mu}(\lambda_c)\nu(\mathcal{E}(\lambda_c)) \nonumber\\
& =
\left(1-\left(1-\frac{r_0}{r_c}\right)^n\right)\sum_{\lambda_c \in \Lambda_c}f_{\theta,\mu}(\lambda_c)\nu_c \nonumber\\
& = 1-\left(1-\frac{r_0}{r_c}\right)^n.
\label{eq:e2}
\end{align}
Relations (\ref{eq:e1})-(\ref{eq:e2}), together with the fact that the first integral in (\ref{eq:e1}) approaches $0$ in the limit of (\ref{lim0}), imply that
\begin{equation*}
\mathbb{P}[U_1\neq U_2] \leq  1-\left(1-\frac{r_0}{r_c}\right)^n + \mathcal{P}_{X_1X_2}(r_0) + o(1).
\end{equation*}
Finally, by applying the above inequality in (\ref{eq:e00}), the conclusion follows.
\end{IEEEproof}

\section{Proofs of Lemmas}
\label{app:lemmas}

\begin{IEEEproof}[Proof of Lemma \ref{d_sup}]
Throughout the proof we will use the fact that $\mu$ is an even integer and, consequently, $\mu\geq 2$, which implies that $\bar{r}_c, \bar{r}_{in}\leq 1/2 \bar{r}_s$.
Using further the fact that $\lambda_{c,i} = \lambda_i + u_i$ and the triangle inequality, we obtain that
\begin{align}
\label{th3_eq1}
\|x^n_i-\hat{x}_{c,i}^n\| & =  \|x^n_i- \lambda_{c,i} + u_i + \lambda_i - \hat{x}_{c,i}^n\| \nonumber \\
&\leq  \|x^n_{c,i}-\lambda_{c,i}\| +\|u_i\| + \|\lambda_i- \hat{x}^n_{c,i}\| \nonumber \\
& \leq  \bar{r}_c +\bar{r}_{in} +  \|\lambda_i- \hat{x}^n_{c,i}\|\nonumber \\
& \leq \bar{r}_s + \|\lambda_i- \hat{x}^n_{c,i}\|.
\end{align}
If condition (\ref{dec_cond}) is violated, then $\hat{x}^n_{c,i}=\lambda_{s,i}$. Thus, we have
\begin{eqnarray*}
\|\lambda_i- \hat{x}^n_{c,i}\|
=  \|\lambda_i- \lambda_{s,i}\| \leq \bar{r}(\beta_i^{-1}({\bf{0}})) \leq (4+\mu/2)\bar{r}_s,
\end{eqnarray*}
where the last inequality is from (\ref{bar_r}). The above relations, together with (\ref{th3_eq1}), imply that
\begin{align*}
\|x_i^n-\hat{x}_{c,i}^n \|\leq (5+\mu/2)\bar{r}_s\leq 3c\bar{r}_s,
\end{align*}
proving that the claim holds when (\ref{dec_cond}) is not true.

Let us assume now that condition (\ref{dec_cond}) is satisfied and that Case 3) holds at the decoder, i.e., $b_1=b_2=1$ and $\tau_1\ne \tau_2$. Thus, $\hat{x}_{c,i}^n=\tilde{\lambda}_i + u_i$, where $\tilde{\lambda}_i$ is given in
(\ref{decoder_step32a}) and (\ref{decoder_step32b}). Then
\begin{align}
\label{th3_eq3}
\|\lambda_i- \hat{x}^n_{c,i}\|
 \leq  \|\lambda_i- \tilde{\lambda}_{i}\| + \|u_i\| \le \|\lambda_i- \tilde{\lambda}_{i}\| + \bar{r}_{in}.
\end{align}
Let us consider now $i=1$. Using (\ref{decoder_step32a}), (\ref{decoder_step32b}) and the triangle inequality, we obtain that
\begin{align}
\label{th3_eq4}
&\|\lambda_1- \tilde{\lambda}_{1}\|\nonumber\\
& \leq  \|\lambda_1- \tilde{\lambda}_{s}\| + \|\tau_1\| + \frac{1}{\mu}\|\lambda_{s,1}-\tilde{\lambda}_{s}\|\nonumber\\
& \leq   \|\lambda_1-\lambda_{s,1}\| + \|\lambda_{s,1}-\tilde{\lambda}_{s}\| + \bar{r}_s + \frac{1}{\mu}\|\lambda_{s,1}-\tilde{\lambda}_{s}\|\nonumber\\
& \leq  (4+\mu/2)\bar{r}_s + \bar{r}_s + \left(1+\frac{1}{\mu}\right)\|\lambda_{s,1}-\tilde{\lambda}_{s}\|,
\end{align}
where the last inequality is based on $\|\lambda_1-\lambda_{s,1}\|\leq \bar{r}(\beta_i^{-1}({\bf{0}}))$ and (\ref{bar_r}).
Using now (\ref{decoder_step31b}) in conjunction with the triangle inequality leads to
\begin{align}
\label{th3_eq5}
\|\lambda_{s,1}-\tilde{\lambda}_{s}\|
& \leq \|\lambda_{s,1}- \tilde{v}\| + (\mu+1)\|\tilde{w}\| \nonumber \\
& \leq \|\lambda_{s,1}- \tilde{v}\| + 2(\mu+1)\bar{r}_s,
\end{align}
where the last inequality follows based on (\ref{tilde_w}) and the fact that
\begin{align*}
\|\tilde{w}\| & \leq \| \hat{w} + \frac{1}{2}(\tau_2-\tau_1) - Q_s(\hat{w} + \frac{1}{2}(\tau_2-\tau_1))\| \\
& \quad  + \|\frac{1}{2}(\tau_2-\tau_1)\| \\
&\leq 2\bar{r}_s.
\end{align*}
Finally, based on (\ref{decoder_step31aa}) and (\ref{decoder_step31ab}), we obtain that
\begin{align}
\label{th3_eq6}
\|\lambda_{s,1}-\tilde{v}\|
& = \left \|\frac{1}{2}(\lambda_{s,1}-\lambda_{s,2}) - \frac{1}{2}\mu\tilde{\lambda} + \left ( 1 + \frac{\mu}{2} \right)\tau_2 - \frac{\mu}{2}\tau_1 \right \|  \nonumber \\
& \leq \frac{1}{2} \|(\lambda_{s,1}-\lambda_{s,2})\| + \frac{1}{2}\mu\|\tilde{\lambda}\| + \left ( 1 + \frac{\mu}{2} \right)\|\tau_2 \| \nonumber \\
& \quad + \frac{\mu}{2}\|\tau_1\|.
\end{align}
Notice that relation (\ref{dec_cond}) implies that
\begin{equation}
\|(\lambda_{s,1}-\lambda_{s,2})\| \leq (10+\mu)\bar{r}_s.
\label{dec_cond2}
\end{equation}
Additionally, from (\ref{delta_lambda}) we obtain that
\begin{align}
\|\tilde{\lambda}\| & \leq \|u_1-u_2\| + \|(u_1-u_2)-Q_{in}(u_1-u_2)\| \nonumber \\
&\leq 2\bar{r}_{in} + \bar{r}_{in} \leq  2\bar{r}_{s}.
\label{delta_lambda2}
\end{align}
Plugging (\ref{dec_cond2}) and (\ref{delta_lambda2}) in (\ref{th3_eq6})
leads to
\begin{align*}
\|\lambda_{s,1}-\tilde{v}\|
 \leq  (5+\mu/2)\bar{r}_s + \bar{r}_{s} + (1+\mu)\bar{r}_s = (7+3\mu/2)\bar{r}_s.
\end{align*}
The above relations and (\ref{th3_eq5})
imply that
\begin{align*}
\|\lambda_{s,1}-\tilde{\lambda}_{s}\|
 \leq   (9+7\mu/2)\bar{r}_s.
\end{align*}
Combining now the above inequality with (\ref{th3_eq1}), (\ref{th3_eq3}) and (\ref{th3_eq4}), we obtain that
\begin{align*}
\|x^n_1-\hat{x}_{c,1}^n\|
& \leq   \bar{r}_s + 1/2\bar{r}_{s} + (5+\mu/2)\bar{r}_s \\
& \quad + \left(1+\frac{1}{\mu}\right)(9+7\mu/2)\bar{r}_s \\
& \leq (24+4\mu)\bar{r}_s
\\ & \leq 16\mu\bar{r}_s,
\end{align*}
which proves the claim. The proof for $i=2$ and for the remaining cases follows along the same lines.
\end{IEEEproof}

\begin{IEEEproof}[Proof of Lemma \ref{probGoTo0}]
Let us fix $i$.
Denote
\begin{align*}
\tilde{\mathcal{C}}(\lambda_{s/2})&\triangleq \{x_i^n \in \mathbb{R}^n: Q_{in}(Q_{c}(x_{i}^{n}))\in \mathcal{C}(\lambda_{s/2})\}, \\
\tilde{\mathcal{C}}&\triangleq \cup_{\lambda_{s/2}\in \Lambda_{s/2}}\tilde{\mathcal{C}}(\lambda_{s/2}).
\end{align*}
A moment of thought reveals that
\begin{align*}
\tilde{\mathcal{C}}(\lambda_{s/2})\subset \left(\lambda_{s/2} + \gamma_1V_{s/2}({\bf{0}})\right ) \setminus \left(\lambda_{s/2} + \gamma_2V_{s/2}({\bf{0}})\right ),
\end{align*}
where $\gamma_1=1 +\frac{\bar{r}_{in}+\bar{r}_c}{{r}_{s/2}}$ and $\gamma_2=\gamma - \frac{\bar{r}_{in}+\bar{r}_c}{{r}_{s/2}}$. The above relation implies that
\begin{equation}
\label{nu_tilde_c}
\nu(\tilde{\mathcal{C}}(\lambda_{s/2}))\leq (\gamma_1^n-\gamma_2^n)\nu(V_{s/2}(\lambda_{s/2})).
\end{equation}
Let
\begin{align*}
\tilde{\mathcal{V}}(\lambda_{s/2})\triangleq \{x_i^n \in \mathbb{R}^n | Q_{in}(Q_{c}(x_{1}^{n}))\in V_{s/2}(\lambda_{s/2})\}.
\end{align*}
Clearly, $\nu(\tilde{\mathcal{V}}(\lambda_{s/2}))=\nu_{s/2}$. The proof of the lemma hinges on the fact that,
under (\ref{lim0}), the pdf of $X_i^n$ can be approximated by a pdf which is uniform over
$ \tilde{\mathcal{V}}_{s/2}(\lambda_{s/2})$. The general idea of the proof is as follows. We have
\begin{align*}
&\mathbb{P}[Q_{in}(Q_{c}(X_{i}^{n}))\in \mathcal{C}(\lambda_{s/2})]\\
&\stackrel{(a)}{\approx}
f_{X_{i}^{n}}(\lambda_{s/2})\nu(\tilde{\mathcal{C}}(\lambda_{s/2}))\\
& \leq f_{X_{i}^{n}}(\lambda_{s/2})\nu(V(\lambda_{s/2}))(\gamma_1^n-\gamma_2^n),
\end{align*}
where the last inequality follows from (\ref{nu_tilde_c}). The above relations lead to
\begin{align*}
& \mathbb{P}[Q_{in}(Q_{c}(X_{i}^{n}))\in \cup_{\lambda_{s/2}\in \Lambda_{s/2}}\mathcal{C}(\lambda_{s/2})]\\
& \qquad \leq \sum_{\lambda_{s/2}\in \Lambda_{s/2}}f_{X_{i}^{n}}(\lambda_{s/2})\nu(V(\lambda_{s/2}))(\gamma_1^n-\gamma_2^n)\\
& \qquad \stackrel{(b)}{\approx} \gamma_1^n-\gamma_2^n,
\end{align*}
where (b) follows from the assumption that the pdf is uniform over $V_{s/2}(\lambda_{s/2})$, thus $\sum\limits_{\lambda_{s/2}\in \Lambda_{s/2}}f_{X_{i}^{n}}(\lambda_{s/2})\nu(V(\lambda_{s/2}))=1$. Finally, it is easy to see that $\gamma_1\rightarrow 1$  and $\gamma_2\rightarrow 1$   under (\ref{lim0}), thus $\lim\limits_{(\ref{lim0})}(\gamma_1^n-\gamma_2^n) =0$.

Next we provide a detailed proof, which includes a rigorous treatment of relations (a) and (b).
Note that the sets $\tilde{\mathcal{V}}(\lambda_{s/2})$ with $\lambda_{s/2}\in \Lambda_{s/2}$ form a partition of $\mathbb{R}^n$. Define a density function $f_{\theta,\mu}:\mathbb{R}^n\rightarrow [0,\infty)$, which is uniform on each set ${\tilde{\mathcal{V}}(\lambda_{s/2})}$, as follows
\begin{equation}
\label{def_sc}
f_{\theta,\mu}(x^n)=\frac{1}{\nu(\tilde{\mathcal{V}}(\lambda_{s/2}))}\int_{\tilde{\mathcal{V}}(\lambda_{s/2})} f_{X_i^n}(y^n)dy^n,\ x^n\in \tilde{\mathcal{V}}(\lambda_{s/2}).
\end{equation}
In view of Lemma \ref{lemma1}, which is stated and proved after the proof of this lemma, we have that $f_{\theta,\mu}(x^n)\rightarrow f_{X_i^n}(x^n)$, for $x^n\in \mathbb{R}^n$, under (\ref{lim0}). Further, we have
\begin{align*}
\mathbb{P}[X_i^n \in \tilde{\mathcal{C}}]&=\int_{\tilde{\mathcal{C}}}\left(f_{X_i^n}(x^n)-f_{\theta,\mu}(x^n)+f_{\theta,\mu}(x^n)\right)dx^n \\
&\leq   \int_{\tilde{\mathcal{C}}}|f_{X_i^n}(x^n)-f_{\theta,\mu}(x^n)|dx^n \\
& \quad +
\sum_{\lambda_{s/2}\in \Lambda_{s/2}}\int_{\tilde{\mathcal{C}}(\lambda_{s/2})}f_{\theta,\mu}(x^n)dx^n.
\end{align*}
Note that
\begin{align*}
& \int_{\tilde{\mathcal{C}}}|f_{X_i^n}(x^n)-f_{\theta,\mu}(x^n)|dx^n \\ & \qquad \qquad \leq
\int_{\mathbb{R}^n}|f_{X_i^n}(x^n)-f_{\theta,\mu}(x^n)|dx^n \rightarrow 0 \mbox{ under (\ref{lim0})},
\end{align*}
where the last relation is valid in view of Scheffe's theorem \cite{scheffe}.
Further, since the density $f_{\theta,\mu}$ is uniform over each $\tilde{\mathcal{V}}(\lambda_{s/2})$ and $\tilde{\mathcal{C}}(\lambda_{s/2})\subset \tilde{\mathcal{V}}(\lambda_{s/2})$, we obtain that
\begin{align*}
&\sum_{\lambda_{s/2}\in \Lambda_{s/2}}\int_{\tilde{\mathcal{C}}(\lambda_{s/2})}f_{\theta,\mu}(x^n)dx^n \\
& \qquad = \sum_{\lambda_{s/2}\in \Lambda_{s/2}} f_{\theta,\mu}(\lambda_{s/2})\nu(\tilde{\mathcal{C}}(\lambda_{s/2}))\nonumber \\
& \qquad \stackrel{(c)}{\leq}
\sum_{\lambda_{s/2}\in \Lambda_{s/2}} f_{\theta,\mu}(\lambda_{s/2})\nu(V_{s/2}(\lambda_{s/2}))(\gamma_1^n-\gamma_2^n)\\
& \qquad =
(\gamma_1^n-\gamma_2^n)\sum_{\lambda_{s/2}\in \Lambda_{s/2}} f_{\theta,\mu}(\lambda_{s/2})\nu(V_{s/2}(\lambda_{s/2}))\\
& \qquad \stackrel{(d)}{=} (\gamma_1^n-\gamma_2^n)\sum_{\lambda_{s/2}\in \Lambda_{s/2}}\int_{\tilde{\mathcal{V}}(\lambda_{s/2})}f_{X_1^n}(y^n)dy^n\\
& \qquad = (\gamma_1^n-\gamma_2^n) \int_{\mathbb{R}^n}f_{X_i^n}(y^n)dy^n = \gamma_1^n-\gamma_2^n,
\end{align*}
where (c) follows from (\ref{nu_tilde_c}), and (d) is based on relation (\ref{def_sc}) and the fact that $\nu(V_{s/2}(\lambda_{s/2}))=\nu(\tilde{\mathcal{V}}(\lambda_{s/2}))$. This observation concludes the proof.
\end{IEEEproof}

\begin{lemma}
\label{lemma1}
Let $\Lambda$ be a lattice and $\sigma>0$ a scale factor. Let $\mathcal{C}_{\sigma}$ be a measurable fundamental cell of the scaled lattice $\sigma\Lambda$ such that  $\lim\limits_{\sigma\rightarrow 0}\bar{r}(\mathcal{C}_{\sigma})=0$.
Let $f:\mathbb{R}^n\rightarrow [0,\infty)$ be a continuous density function. For each $\sigma$, define the function
$f_{\sigma}:\mathbb{R}^n\rightarrow [0,\infty)$ as follows. For each $\lambda_{\sigma} \in \sigma \Lambda$ and $x^n \in \lambda_{\sigma} + \mathcal{C}_{\sigma}$, let
\begin{equation}
f_{\sigma}(x^n)\triangleq \frac{1}{\nu(\mathcal{C}_{\sigma})}\int_{\lambda_{\sigma} + \mathcal{C}_{\sigma}}f(y^n)dy^n.
\end{equation}
Then for every $x^n\in \mathbb{R}^n$,
\begin{equation}
\label{lim_lemma1}
\lim_{\sigma\rightarrow 0}f_{\sigma}(x^n)=f(x^n).
\end{equation}
\end{lemma}
\begin{IEEEproof}
Let us fix $x^n\in \mathbb{R}^n$ and let  $\lambda_{\sigma} \in \sigma\Lambda$ such that $x^n \in \lambda_{\sigma} + \mathcal{C}_{\sigma}$. Then
\begin{align}
\label{eq:2001}
|f_{\sigma}(x^n)-f(x^n)|&\leq
\frac{1}{\nu(\mathcal{C}_{\sigma})}\int_{\lambda_{\sigma} + \mathcal{C}_{\sigma}}|f(y^n)-f(x^n)|dy^n   \nonumber \\
& \leq \max_{y^n\in \overline{\lambda_{\sigma} + \mathcal{C}_{\sigma}}}|f(y^n)-f(x^n)| \nonumber \\ & \leq \max_{y^n\in \overline{x^n + \mathcal{B}_{2\bar{r}(\mathcal{C}_{\sigma})}}}|f(y^n)-f(x^n)|.
\end{align}
Since $f$ is continuous and  the set $x^n + \mathcal{B}_{2\bar{r}(\mathcal{C}_{\sigma})}$ is a neighborhood of $x^n$ with diameter approaching $0$ as $\sigma\rightarrow 0$, it  further follows that
\begin{eqnarray}
\label{eq:2002}
\lim_{\sigma\rightarrow 0}\max_{y^n\in \overline{x^n + \mathcal{B}_{2\bar{r}(\mathcal{C}_{\sigma})}}}|f(y^n)-f(x^n)|=0.
\end{eqnarray}
Relations (\ref{eq:2001}) and (\ref{eq:2002}) imply that (\ref{lim_lemma1}) holds.
\end{IEEEproof}

\begin{lemma}
\label{lemma:shift}
For any set $\mathcal{A}\subseteq \mathbb{R}^n$ and any $u \in \mathbb{R}^n$,
\begin{align*}
&\int_{u+\mathcal{A}}\|x^n\|^2 dx^n \\
&= \int_{\mathcal{A}}\|x^n\|^2 dx^n
+ 2\langle\int_{\mathcal{A}}x^n dx^n, u \rangle + \|u\|^2 \nu(\mathcal{A}).
\end{align*}
\end{lemma}
\begin{IEEEproof}
Applying the change of variable $x^n=u+y^n$, we obtain that
\begin{align*}
&\int_{u+\mathcal{A}}\|x^n\|^2 dx^n \\
& = \int_{\mathcal{A}}\|y^n+u\|^2 dy^n \\
& =
\int_{\mathcal{A}}\|y^n\|^2 dy^n + \int_{\mathcal{A}} 2\langle y^n, u \rangle dy^n  + \int_{\mathcal{A}}\|u\|^2 dy^n \\
& = \int_{\mathcal{A}}\|y^n\|^2 dy^n + 2\langle \int_{\mathcal{A}}x^n dx^n, u \rangle  + \|u\|^2 \nu(\mathcal{A}).
\end{align*}
\end{IEEEproof}

\begin{lemma}
\label{sum_squared_norms}
Let $\Lambda_2\subset \Lambda_1$ be two nested lattices in $\mathbb{R}^n$. Let $N_0\triangleq N(\Lambda_2:\Lambda_1)$ and $\mathcal{C}_{\Lambda_2:\Lambda_1}\triangleq
\cup_{\lambda_1\in V_{\Lambda_2}({\bf{0}})\cap \Lambda_1}V_{\Lambda_1} (\lambda_1)$.
Then
\begin{equation*}
\label{nested_sum}
\sum_{\lambda_1\in V_{\Lambda_2}({\bf{0}})\cap \Lambda_1}
\|\lambda_1 \|^2 = nN_0\left(G(\mathcal{C}_{\Lambda_2:\Lambda_1})\nu_{\Lambda_2}^{\frac{2}{n}}- G_{\Lambda_1}\nu_{\Lambda_1}^{\frac{2}{n}}\right).
\end{equation*}
\end{lemma}
\begin{IEEEproof}
It can be easily seen that $\mathcal{C}_{\Lambda_2:\Lambda_1}$ is a fundamental region of the lattice $\Lambda_2$, thus  $\nu(\mathcal{C}_{\Lambda_2:\Lambda_1})=\nu(\Lambda_2)=N_0 \nu(\Lambda_1)$. Invoking further the definition of  $G(\mathcal{C}_{\Lambda_2:\Lambda_1})$ gives
\begin{equation*}
nN_0 G(\mathcal{C}_{\Lambda_2:\Lambda_1})\nu_{\Lambda_2}^{\frac{2}{n}} = \frac{1}{\nu_{\Lambda_1}}
\int_{\mathcal{C}_{\Lambda_2:\Lambda_1}} \| x^n \|^2 dx^n.
\end{equation*}
Using the fact that $V_{\Lambda_1}(\lambda_1)=\lambda_1 + V_{\Lambda_1}({\mathbf{0}})$, we obtain that
\begin{align*}
& \frac{1}{\nu_{\Lambda_1}} \int_{\mathcal{C}_{\Lambda_2:\Lambda_1}} \| x^n \|^2 dx^n \\
&\qquad = \frac{1}{\nu_{\Lambda_1}} \sum_{\lambda_1\in V_{\Lambda_2}({\bf{0}})\cap \Lambda_1}
\int_{\lambda_1 + V_{\Lambda_1}({\mathbf{0}})} \| x^n \|^2 dx^n\\
& \qquad \stackrel{(a)}{=}  \frac{1}{\nu_{\Lambda_1}} \sum_{\lambda_1\in V_{\Lambda_2}({\bf{0}})\cap \Lambda_1}
\left ( \int_{V_{\Lambda_1}({\bf{0}})} \| x^n\|^2 dx^n \right. \\
& \qquad \qquad \left. + 2\left \langle \int_{V_{\Lambda_1}({\mathbf{0}})}  x^n dx^n, \lambda_1 \right \rangle  +
\| \lambda_1 \|^2 \nu_{\Lambda_1} \right )\\
& \qquad \stackrel{(b)}{=}  \frac{N_0}{\nu_{\Lambda_1}}
 \int_{V_{\Lambda_1}({\bf{0}})} \| x^n\|^2 dx^n +
\sum_{\lambda_1\in V_{\Lambda_2}({\bf{0}})\cap \Lambda_1} \| \lambda_1 \|^2 \\
& \qquad \stackrel{(c)}{=}  nN_0G_{\Lambda_1}\nu_{\Lambda_1}^{\frac{2}{n}} +
\sum_{\lambda_1\in V_{\Lambda_2}({\bf{0}})\cap \Lambda_1} \| \lambda_1 \|^2.
\end{align*}
Here (a) is based on Lemma \ref{lemma:shift}; moreover, (b) uses the fact that
$\int_{V_{\Lambda_1}({\mathbf{0}})}  x^n dx^n={\mathbf{0}}$ and $|V_{\Lambda_2}({\bf{0}})\cap \Lambda_1|=N_0$, while (c) is based on the definition of $G_{\Lambda_1}$. Now the claim follows.
\end{IEEEproof}

\begin{lemma}
\label{shape}
Consider two nested lattices $\Lambda_{2,0}\subset \Lambda_{1,0}$ and the scale coefficients $\omega_1, \omega_2$ such that lattices $\Lambda_2=\omega_2 \Lambda_{2,0}$ and $\Lambda_1=\omega_1 \Lambda_{1,0}$ are still nested. Let $N_0\triangleq N(\Lambda_2:\Lambda_1)$ and $\mathcal{C}_{\Lambda_2:\Lambda_1}\triangleq
\cup_{\lambda_1\in V_{\Lambda_2}({\bf{0}})\cap \Lambda_1}V_{\Lambda_1} (\lambda_1)$. Then
\begin{equation*}
\lim_{\frac{\omega_2}{\omega_1}\rightarrow\infty}G(\mathcal{C}_{\Lambda_2:\Lambda_1})= G_{\Lambda_2}.
\end{equation*}
\end{lemma}
\begin{IEEEproof}
Since the lattices $\Lambda_{2}$ and $\Lambda_{1}$ are scaled by different scale factors, the value $G(\mathcal{C}_{\Lambda_{2}:\Lambda_{1}})$ is not constant. On the other hand, $G_{\Lambda_2}$ is constant. Notice further that  the set  $\mathcal{C}_{\Lambda_{2}:\Lambda_{1}}$ is a fundamental region of the lattice $\Lambda_{2}$, thus its volume equals $\nu_{\Lambda_2}$. Then the following holds
\begin{align*}
G(\mathcal{C}_{\Lambda_{2}:\Lambda_{1}})- G_{\Lambda_2}
&= \frac{1}{n\nu_{\Lambda_{2}}^{1+\frac{2}{n}}}\left (
\int_{\mathcal{C}_{\Lambda_{2}:\Lambda_{1}}}\|x^n\|^2 dx^n \right.\\
& \qquad \qquad \quad  \left.-
\int_{V_{\Lambda_{2}}({\bf{0}})}\|x^n\|^2 dx^n\right ).
\end{align*}
For simplicity, let us denote $\mathcal{A}=\mathcal{C}_{\Lambda_{2}:\Lambda_{1}}$, $\mathcal{B}=V_{\Lambda_{2}}({\bf{0}})$ and $\Delta \nu=\nu(\mathcal{A})- \nu(\mathcal{A}\cap \mathcal{B})$. Since $\nu(\mathcal{A})=\nu(\mathcal{B})$, it follows that $\Delta \nu=\nu(\mathcal{B})- \nu(\mathcal{A}\cap \mathcal{B})$. Then we obtain  that
\begin{align*}
|G(\mathcal{A})- G(\mathcal{B})|
&= \frac{1}{n\nu_{\Lambda_{2}}^{1+\frac{2}{n}}}\left |
\int_{\mathcal{A}\setminus \mathcal{A}\cap \mathcal{B}}\|x^n\|^2 dx^n \right.\\
& \qquad \qquad \quad \left. -
\int_{\mathcal{B}\setminus \mathcal{A}\cap \mathcal{B}}\|x^n\|^2 dx^n\right | \nonumber \\
&\leq  \frac{1}{n\nu_{\Lambda_{2}}^{1+\frac{2}{n}}} \left ( \bar{r}(\mathcal{A})^2 \Delta \nu + \bar{r}(\mathcal{B})^2 \Delta \nu \right ) \nonumber \\
&\leq  \frac{\Delta \nu}{n\nu_{\Lambda_{2}}^{1+\frac{2}{n}}}
\left ( (\bar{r}_{\Lambda_{2}}+\bar{r}_{\Lambda_{1}})^2  + \bar{r}_{\Lambda_{2}}^2 \right )
\nonumber \\
&\leq \frac{5\bar{r}_{\Lambda_{2}}^2\Delta \nu}{n\nu_{\Lambda_2}^{1+\frac{2}{n}}} =
\frac{5\omega_2^2\bar{r}_{\Lambda_{2,0}}^2\Delta \nu}{n\omega_2^2\nu_{\Lambda_{2,0}}^{\frac{2}{n}}\nu_{\Lambda_{2}}} = \frac{5\bar{r}_{\Lambda_{2,0}}^2}{4n\nu_{\Lambda_{2,0}}^{\frac{2}{n}}}
\frac{\Delta \nu}{\nu_{\Lambda_{2}}}.
\end{align*}
According to the above relations, in order to prove the claim of the lemma, it is sufficient to show that $\lim\limits_{\frac{\omega_2}{\omega_1}\rightarrow\infty} \frac{\Delta \nu}{\nu_{\Lambda_{2}}}=0$, which is equivalent to
\begin{equation}
\label{delta_nu}
\lim_{\frac{\omega_2}{\omega_1}\rightarrow\infty} \frac{\nu(\mathcal{A}\cap \mathcal{B})}{\nu_{\Lambda_{2}}}=1.
\end{equation}
It is easy to see that, for any point $x^n \in V_{\Lambda_{2}}({\bf{0}})$ which is at a distance larger than $\bar{r}_{\Lambda_{1}}$ from the boundary of $V_{\Lambda_{2}}({\bf{0}})$, we have $Q_{\Lambda_{1}}(x^n)\in V_{\Lambda_{2}}({\bf{0}})$, thus $x^n \in \mathcal{A}$. This observation implies that the interior of the set $\gamma V_{\Lambda_{2}}({\bf{0}})$ is included in $\mathcal{A}\cap \mathcal{B}$, where
\begin{align*}
\gamma = 1-\frac{\bar{r}_{\Lambda_{1}}}{r_{\Lambda_{2}}}=1-\frac{\omega_1}{\omega_2}\frac{\bar{r}_{\Lambda_{1,0}}}{{r}_{\Lambda_{2,0}}}.
\end{align*}
Then we have
$\gamma^n\leq \frac{\nu(\mathcal{A}\cap \mathcal{B})}{\nu_{\Lambda_{2}}}\leq 1$, which implies that (\ref{delta_nu}) holds. With this the proof is completed.
\end{IEEEproof}

\section*{Acknowledgment}
The authors would like to thank the Associate Editor and the anonymous reviewers for their valuable comments and suggestions, which helped improve the quality of the work.


\ifCLASSOPTIONcaptionsoff
  \newpage
\fi

\end{document}